\documentclass[12pt]{iopart}

\usepackage{iopams}
\usepackage{amstext}
\usepackage{mathabx}
\usepackage{graphicx}
\usepackage{verbatim} 
\usepackage{color}

\newcommand{\aop}{\hat{a}}

\newcommand{\bopd}{\hat{b}^\dagger}
\newcommand{\bop}{\hat{b}}

\newcommand{\atl}{\tilde{a}(\omega)}
\newcommand{\atlss}[1]{\tilde{a}_{\text{#1}}(\omega)}

\newcommand{\btl}{\tilde{b}(\omega)}
\newcommand{\btlss}[1]{\tilde{b}_{\text{#1}}(\omega)}

\newcommand{\btlssd}[1]{\tilde{b}^\dagger_{\text{#1}}(\omega)}
\newcommand{\atlssd}[1]{\tilde{a}^\dagger_{\text{#1}}(\omega)}
\newcommand{\btlssp}[1]{\tilde{b}_{\text{#1}}(\omega^\prime)}
\newcommand{\atlssp}[1]{\tilde{a}_{\text{#1}}(\omega^\prime)}

\newcommand{\m}{\mathbf}

\newcommand{\D}{\Delta}

\newcommand{\twotwomat}[4]{\left(
\begin{array}{cc}
{#1} & {#2} \\
{#3} & {#4} \\
\end{array}
\right) }

\newcommand{\twovec}[2]{\left(
\begin{array}{c}
{#1} \\
{#2} \\
\end{array}
\right) }

\def\be{\begin{equation}}
\def\ee{\end{equation}}
\def\bea{\begin{eqnarray}}
\def\eea{\end{eqnarray}}
\newcommand{\ket}[1]{\mbox{$|#1\rangle$}}
\newcommand{\bra}[1]{\mbox{$\langle#1|$}}
\newcommand{\avg}[1]{\mbox{$\langle#1\rangle$}}

\newcommand{\opdagger}[2]{\mbox{$\hat{#1}_{#2}^{\dagger}$}}
\newcommand{\op}[2]{\mbox{$\hat{#1}_{#2}$}}

\begin{document}

\title[Proposal for an Optomechanical Traveling Wave Phonon-Photon Translator]{Proposal for an Optomechanical Traveling Wave Phonon-Photon Translator}

\author{Amir H. Safavi-Naeini, Oskar Painter}

\address{Thomas J. Watson, Sr., Laboratory of Applied Physics, California Institute of Technology, Pasadena, CA 91125}
\ead{safavi@caltech.edu, opainter@caltech.edu}
\begin{abstract}
In this article we describe a general optomechanical system for converting photons to phonons in an efficient, and reversible manner.  We analyze classically and quantum mechanically the conversion process and proceed to a more concrete description of a phonon-photon translator formed from coupled photonic and phononic crystal planar circuits. Applications of the phonon-photon translator to RF-microwave photonics and circuit QED, including proposals utilizing this system for optical wavelength conversion, long-lived quantum memory and state transfer from optical to superconducting qubits are considered. 
\end{abstract}

\maketitle

\section{Introduction}

Classical and quantum information processing network architectures utilize light (optical photons) for the transmission of information over extended distances, ranging from hundreds of meters to hundreds of kilometers~\cite{Agrawal2002,Duan2001}. The utility of optical photons stems from their weak interaction with the environment, large bandwidth of transmission, and resiliency to thermal noise due to their high frequency ($\sim 200~\text{THz}$).  Acoustic excitations (phonons), though limited in terms of bandwidth and their ability to transmit information farther than a few millimeters, can be delayed and stored for significantly longer times and can interact resonantly with RF-microwave electronic systems~\cite{Pozar2004}.  This complimentary nature of photons and phonons suggests hybrid phononic-photonic systems as a fruitful avenue of research, where a new class of \emph{optomechanical} circuitry could be made to perform a range of tasks out of reach of purely photonic and phononic systems. A building block of such a hybrid architecture would be elements coherently interfacing optical and acoustic circuits. The optomechanical translator we propose in this paper acts as a chip-scale \emph{transparent, coherent interface} between phonons and photons and fulfills a key requirement in such a program.   


In the quantum realm, systems involving optical, superconducting, spin or charge qubits coupled to mechanical degrees of freedom~\cite{Cleland2004,Geller2005,rabl-2009,wallquist-2009-137,Chang2010,Stannigel2010} have been explored.  The recent demonstration of coherent coupling between a superconducting qubit and a mechanical resonance by O'Connell, et al.~\cite{OConnell2010}, has provided an experimental backing for this vision and is the latest testament to the versatility of mechanics as a connecting element in hybrid quantum systems.  In the specific case of phonon-photon state transfer, systems involving trapped atoms, ions, nanospheres~\cite{Parkins1999,Massoni2000,Orszag2002,Rodrigues2006,chang_cavity_2010}, and mechanically compliant optical cavity structures~\cite{Zhang2003} have all been considered. In these past studies, the state of an incoming light field is usually mapped onto the motional state of an atom, ion, or macroscopic mirror, through an exact timing of control pulses, turning on and off the interaction between the light and mechanical motion in a precise way. The ability to simultaneously implement phononic and photonic waveguides in optomechanical crystal (OMC) structures~\cite{Eichenfield2009} opens up the opportunity to implement a~\textit{traveling-wave} phonon-photon translator (PPT).  Such a device, operating continuously, connects acoustic and optical waves to each other in a symmetric manner, and allows for on-the-fly conversion between phonons and photons without having to precisely time the information and control pulses. In effect, the problem of engineering control pulses is converted into a problem of engineering coupling rates.


Our proposal for a PPT is motivated strongly by recent work~\cite{Kippenberg2008,Favero2009} on radiation pressure effects in micro- and nano-scale mechanical systems~\cite{Chan2009,Eichenfield2009-zipper,Li2008,Lin2009,Wiederhecker2009,Roels2009,Safavi-Naeini2010a}.  Furthermore, the concrete realization of a PPT is aided by the considerable advances made in the last decade in the theory, design and engineering of thin-film artificial quasi-2D (patterned membrane) crystal structures containing photonic~\cite{Painter1999,Painter1999a,Vuckovic2001,Takano2006,Song2005a,Notomi2005} and phononic~\cite{Olsson2009,Sanchez-Perez1998,Vasseur2007,Gu2006,Mohammadi2008a} ``band gaps''. Such systems promise unprecedented control over photons and phonons, and have been separately subject to extensive investigation. Their unification, in the form of OMCs which possesses a simultaneous phononic and photonic bandgap~\cite{Eichenfield2009,Eichenfield2009a,Safavi-Naeini2010,Pennec2010,Mohammadi2010}, and in which the interaction between the photons and phonons can be controlled, promises to further expand the capabilities of both photonic and phononic architectures and forms the basis of the proposed PPT implementation.

The outline of this paper is as follows.  In Sections~\ref{sec:outline} and~\ref{sec:analysis} we introduce and study the PPT system as an abstraction, at first classically and then quantum mechanically. After introducing the basic system, its properties and its scattering matrix, we study the effects of quantum and classical noise on device operation. In Section~\ref{sec:implementation} we design and simulate a possible physical implementation of the system, utilizing recent results in simultaneous phononic-photonic bandgap materials~\cite{Safavi-Naeini2010}. Finally, in Section~\ref{sec:applications}, we demonstrate a few possible applications of the PPT. Focusing first on ``classical'' applications, we evaluate the performance of the PPT when used for the implementation of an optical delay line and wavelength converter. Finally, we show how such a system could be used in theory to do high fidelity quantum state transfer between optical and superconducting qubits.

\section{Outline of Proposed System} \label{sec:outline}

\begin{figure}[htbp]
\begin{center}
\scalebox{0.8}{\includegraphics{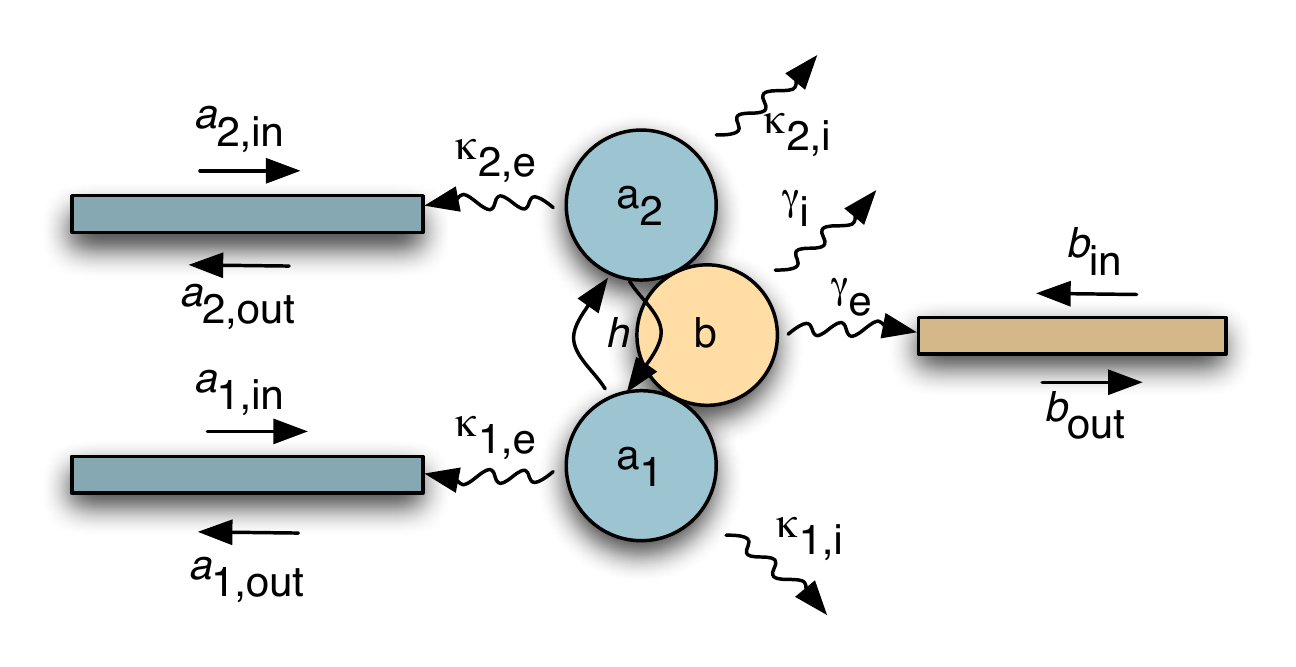}}
\caption{Full system diagram. Circles represent resonant modes, while rectangles represent waveguides. Blue is for photonics, beige is for phononics, a color-scheme followed in other parts of the paper. The coupling $h$ between the two optical modes is modulated by the intervening phonon resonance.}
\label{fig:full_system_diagram}
\end{center}
\end{figure}

The proposed PPT system, shown in Fig.~\ref{fig:full_system_diagram}, consists of a localized mechanical resonance ($b$) which couples the two optical resonances ($a_1$,$a_2$) of an optomechanical cavity via radiation pressure.  External coupling to and from the mechanical resonance is provided by an acoustic waveguide, while each of the optical resonances are coupled to via separate optical waveguides.  Multi-optical-mode optomechanical systems have been proposed and experimentally studied previously in the context of enhancing quantum back-action, reduced lasing threshold, and parametric instabilities~\cite{Braginsky2001,Zou2002,Ju2006,Tomes2009,Grudinin2009a,Dobrindt2010,Grudinin2010}.  Here we use a two-moded optical cavity as it allows for the spatial filtering and separation of signal and pump optical beams while reducing the quantum noise in the system, as is explained below.  

A general description of the radiation-pressure-coupling of the mechanical and optical degrees of freedom in such a structure is as follows.  For each of the two high-$Q$ optical resonances of the cavity we associate an annihilation operator $\aop_k$ and a frequency $\omega_k$ ($k=1,2$).  Geometric deformation of the optomechanical cavity parameterized by $x$, changes the frequencies of the optical modes by $g_k(x)$.  The deformation, due to the localized mechanical resonance with annihilation operator $\bop$ and frequency $\Omega$, can be quantized and given by $\hat{x} = x_\text{ZPF} (\bop + \bopd)$.  There is also a coupling between the two optical cavity modes given by $h(x)$, where for resonant intermodal mechanical coupling the cavity structure must be engineered such that $\Omega = \omega_2 - \omega_1$.  In a traveling wave PPT device consisting of the two optical cavity resonances and a single mechanical resonance, the lower frequency cavity mode ($a_1$ in this case) is used as a 'pump' cavity which enables the inter-conversion of phonons in the mechanical resonance ($b$) to photons in the second, higher frequency, optical cavity mode ($a_2$) through a two-photon process in which pump photons are either absorbed or emitted as needed.  The 'signals' representing the phonon and photon quanta to be exchanged will thus be contained in $b$ and $a_1$, respectively.  

As described, the Hamiltonian of this system is,
\begin{eqnarray} 
 \hat{H} &=& \hbar \omega_1 \hat{a}^\dagger_1 \hat{a}_1 + \hbar \omega_2 \hat{a}^\dagger_2 \hat{a}_2 + \hbar \Omega \hat{b}^\dagger \hat{b} + \hbar(g_1 (\hat{x})\hat{a}^\dagger_1 \hat{a}_1 +g_2 (\hat{x}) \hat{a}^\dagger_2 \hat{a}_2)\nonumber\\ 
&&	+\hbar h (\hat{x})(\hat{a}^\dagger_2 \hat{a}_1 + \hat{a}^\dagger_1 \hat{a}_2) + i \sqrt{2\kappa_{1,e}} E_{\text{pump}} (e^{-i\omega_L t}\hat{a}^\dagger_1 + e^{i\omega_Lt} \hat{a}_1),\label{eqn:full_total_Hamiltonian}
\end{eqnarray}

\noindent where we have added a classical optical pumping term of electric field amplitude $E_{\text{pump}}$ and frequency $\omega_L$.  Optical pumping is performed through one of the optical waveguides with (field) coupling rate to the $a_1$ cavity resonance given by $\kappa_{1,e}$.  In addition to the waveguide loading of each optical resonance ($\kappa_{k,e}$), the total optical loss rate of each cavity mode includes an intrinsic component ($\kappa_{k,i}$) of field decay due to radiation, scattering, and absorption.  Similarly for the mechanical resonance, we have a field decay rate given by $\gamma = \gamma_e + \gamma_i$ which is a combination of waveguide loading and intrinsic losses.  The constant parts of both $h(\hat{x})$ and $g_k(\hat{x})$ ($h(0)$, $g_k(0)$) can be eliminated by a change of basis and are thus taken to be zero. As discussed below, it is advantageous to choose a cavity structure symmetry in which $g_k(\hat{x}) = 0$ up to linear order in $\hat{x}$. In fact, we can generally assume that the mechanical displacements are small enough to make the linear order the only important term in the interaction.  Assuming then a properly chosen cavity symmetry,
\begin{eqnarray}
g_k(\hat{x}) = g\cdot(\bop + \bopd) = 0\qquad\mbox{and}\qquad
h(\hat{x}) = h\cdot(\bop + \bopd),\nonumber
\end{eqnarray}
which yields a simplified Hamiltonian
\begin{eqnarray} 
 \hat{H} &=& \hbar \omega_1 \hat{a}^\dagger_1 \hat{a}_1 + \hbar \omega_2 \hat{a}^\dagger_2 \hat{a}_2 + \hbar \Omega \hat{b}^\dagger \hat{b}+ \hbar h (\hat{b} + \hat{b}^\dagger)(\hat{a}^\dagger_2 \hat{a}_1 + \hat{a}^\dagger_1 \hat{a}_2)\nonumber\\
&&+ i \sqrt{2\kappa_{1,e}} E_{\text{pump}} (e^{-i\omega_L t}\hat{a}^\dagger_1 + e^{i\omega_Lt} \hat{a}_1).\label{eqn:total_Hamiltonian}
\end{eqnarray}

\begin{figure}[htbp]
\begin{center}
\scalebox{0.9}{\includegraphics{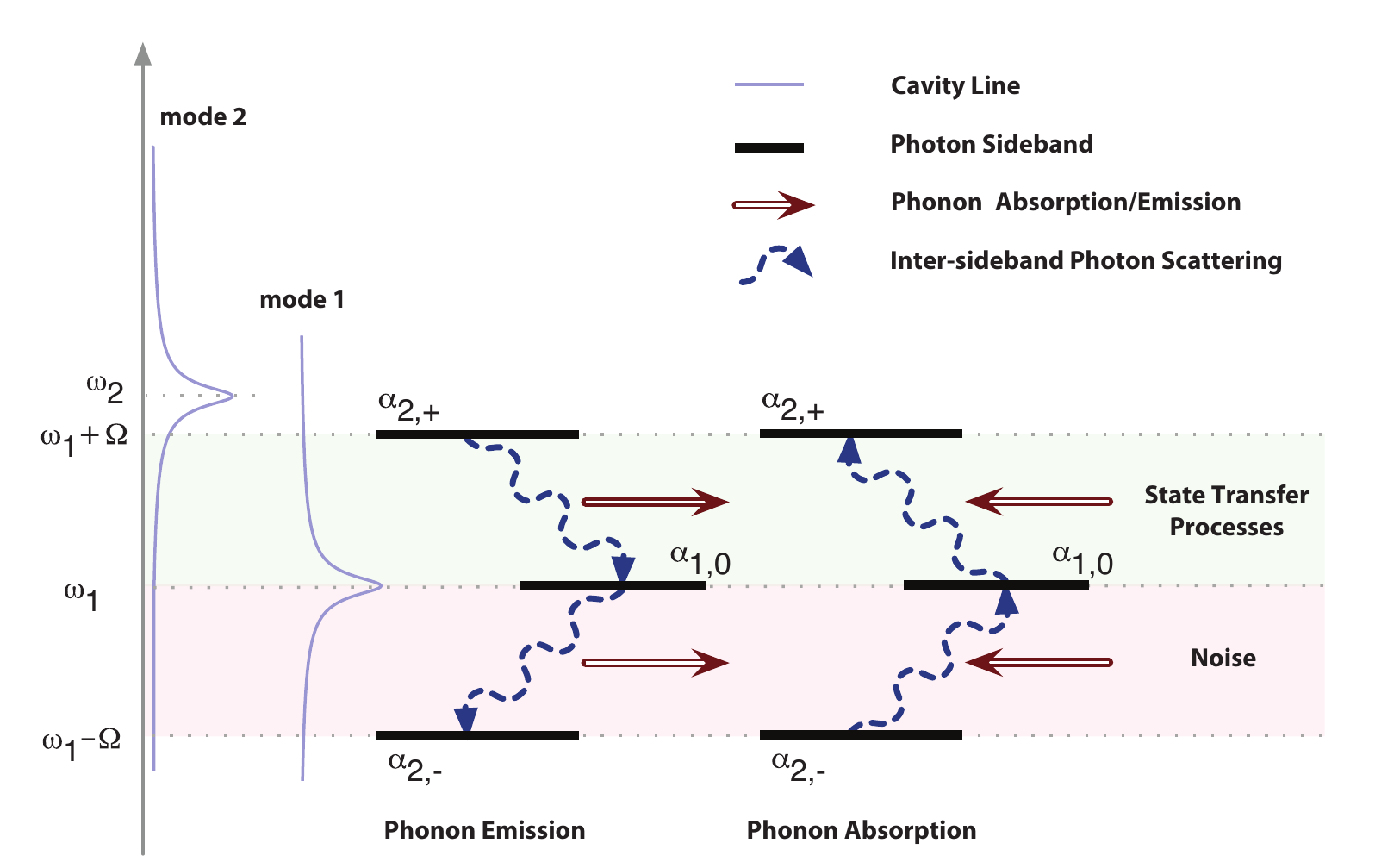}}
\caption{Optical sidebands and processes for phonon-photon translation. The optical pump is located on sideband $\alpha_{1,0}$, on resonance with the first cavity mode at frequency $\omega_1$. There are three relevant optical sidebands to consider in the translation process, $\alpha_{2,+}$, $\alpha_{1,0}$ and $\alpha_{2,-}$. The inter-sideband photon scattering gives rise to phonon emission and absorption. The state transfer occurs through scattering between sidebands $\alpha_{2,+}$ and $\alpha_{1,0}$, whereas inter-sideband scattering between $\alpha_{1,0}$ and $\alpha_{2,-}$ can be thought of as phonon noise. Note that for the $g=0$ case considered here there are no sidebands at $\omega_1\pm\Omega$ for cavity mode $a_1$ and no sidebands at $\omega_1$ for mode $a_2$.}
\label{fig:doublecav_sidebands}
\end{center}
\end{figure}

Treating the system classically and approximately, we can write each intracavity photon and phonon amplitude, and their inputs (see Fig.~\ref{fig:doublecav_sidebands}) as a Fourier decomposition of a few relevant sidebands:
\begin{eqnarray}
a_1(t) &=& \alpha_{1,0}e^{-i\omega_1 t}  + \alpha_{1,+} e^{-i(\omega_1+\Omega) t}+ \alpha_{1,-} e^{-i(\omega_1-\Omega) t} \\
a_2(t) &=& \alpha_{2,0}e^{-i\omega_1 t} + \alpha_{2,+} e^{-i(\omega_1+\Omega)  t} + \alpha_{2,-} e^{-i(\omega_1-\Omega) t} \\
b(t) &=& \beta_{0} + \beta_{+} e^{-i\Omega t}\\
b_\text{in}(t) &=& \beta_{\text{in},+} e^{-i\Omega t}\\
a_\text{2,in}(t) &=& \alpha_{\text{in},+} e^{-i(\omega_1+\Omega) t}\\
b_\text{out}(t) &=&\beta_{\text{out},+} e^{-i\Omega t}\\ 
a_\text{2,out}(t) &=&\alpha_{\text{out},0}e^{-i\omega_1 t}  + \alpha_{\text{out},+} e^{-i(\omega_1+\Omega)  t}
	+ \alpha_{\text{out},-}e^{-i(\omega_1-\Omega) t}.
\end{eqnarray}
The equations of motion arrived at from the system Hamiltonian (presented generally in the following section) then become algebraic relations between the $\alpha$ and $\beta$ sideband amplitudes. By ignoring the self-coupling term ($g=0$), pumping on-resonance with cavity mode $a_1$ ($\omega_L = \omega_1$), and engineering the optical cavity mode splitting for mechanical resonance ($\omega_2 = \omega_1 + \Omega$), we arrive at classical sideband amplitudes,
\begin{eqnarray}
 \alpha_{1,+}&=& \alpha_{1,-} = \alpha_{2,0} = \beta_{0} = 0,\\ 
 \alpha_{1,0} &=& \frac{\sqrt{2 \kappa_e}}{\kappa} E_{\text{pump}} + O(h\alpha_{k,\pm} \beta_+), \\
 \alpha_{2,+} &=& -\frac{i h \alpha_{1,0}}{\kappa}\beta_{+} - \frac{\sqrt{2 \kappa_e}}{\kappa} \alpha_{\text{in},+},\\
 \alpha_{2,-} &=& -\frac{i h \alpha_{1,0}}{\kappa+2 i \Omega}\beta_{+}^\ast,\\
 \beta_{+} &=& -\frac{i h \alpha_{1,0}^\ast}{\gamma}\alpha_{2,+} - \frac{\sqrt{2\gamma_e}}{\gamma} \beta_{\text{in},+} -\frac{i h \alpha_{1,0}}{\gamma}\alpha_{2,-}^\ast.
\end{eqnarray}

From here we see that the central sideband amplitude of cavity mode $a_1$, $\alpha_{1,0}$, is proportional to the sum of a term containing the pump field $E_\text{pump}$ and terms containing products of the optical and mechanical sideband amplitudes. By increasing $E_\text{pump}$ the effect of the other sidebands on the pump resonance amplitude can be made negligible, and we assume here and elsewhere in this work that the pump sideband is generally left unaffected by the dynamics of the rest of the system. As desired the optical sideband which contains mechanical information is $\alpha_{2,+}$ since it is the only sideband directly proportional to $\beta_+$. The constant of proportionality between these two terms is seen to contain both $h$ and $\alpha_{1,0}$, demonstrating the role of the pump beam in the conversion process. Since coherent information transfer between the optics and mechanics is occurring between $\beta_+$ and $\alpha_{2,+}$, it is desirable to remove the effects of the lower energy photonic sideband, $\alpha_{2,-}$. This sideband can be made significantly smaller in magnitude than $\alpha_{2,+}$ in the sideband-resolved regime where $\Omega \gg \kappa$. A convenient way to visualize all of the processes in the system is shown in  Figure~\ref{fig:doublecav_sidebands} where the photonic sideband amplitudes $\alpha_{2,\pm}$ and $\alpha_{1,0}$ are represented as ``energy levels'', with transitions between them being due to the emission and absorption of phonons. 

From this approximate analysis it is clearly suggested that in a sideband resolved optomechanical system, a state-transfer process is possible between the phononic and photonic resonances, and the process is controlled by a pump beam~\cite{Zhang2003}. A more in-depth study of the system dynamics required to understand how such processes may be used for traveling wave phonon-photon conversion, and a full investigation of all relevant noise sources required to understand the applicability of such a system to quantum information, is carried out in the following section.

\section{Analysis} \label{sec:analysis}

A detailed treatment of the operation of a traveling phonon-photon translator is carried out in this section.  At first, the dynamics of the system are simplified while still taking into account the noise processes related to the sideband $\alpha_{2,-}$. In this way one is left with an effective `beam-splitter' Hamiltonian, which describes the coherent interaction between the optics and mechanics, while the aforementioned noise processes are accounted for through an effective increase in the thermal bath temperature. This is followed by a treatment of the traveling-wave problem through a scattering matrix formulation, which provides insight into the role of the intracavity pump photon number ($|\alpha_{1,0}|^2$) and optimizing the state transfer efficiency.

\subsection{Simplified Dynamics of the System}
Starting from the Hamiltonian in equation~(\ref{eqn:total_Hamiltonian}), a set of Heisenberg-Langevin Equations can be written down,~\cite{GardinerZoller}:

\begin{eqnarray}
\dot{\hat{a}}_1 &=& - (\kappa_1 + i \omega_1) \hat{a}_1
- i h (\hat{b}+\hat{b}^\dagger) \hat{a}_2
 + \sqrt{2\kappa_{1,e}} E_{\text{pump}} e^{-i\omega_L t}  - \sqrt{2\kappa_1} \hat{a}^\prime_\text{1,in}, \\
\dot{\hat{a}}_2 &=&  - (\kappa_2 + i \omega_2 ) \hat{a}_2 - 
i h (\hat{b}+\hat{b}^\dagger) \hat{a}_1
 - \sqrt{2\kappa_2} \hat{a}^\prime_\text{2,in}, \\
\dot{\hat{b}} &=& -i \Omega \hat{b} -i h (\hat{a}^\dagger_2 \hat{a}_1 + \hat{a}^\dagger_1 \hat{a}_2) 
- \gamma \hat{b} + \gamma \hat{b}^\dagger - \sqrt{2\gamma} \hat{b}^\prime_\text{in}.
\end{eqnarray}
The input coupling terms as written above include both external waveguide coupling and intrinsic coupling due to lossy channels (see Fig.~\ref{fig:full_system_diagram}).  Separated, the intrinsic (with subscript $i$) and extrinsic (with no subscript) components look as follows,
\begin{eqnarray}
\sqrt{2\kappa_k} \hat{a}^\prime_\text{k,in} &=& \sqrt{2\kappa_{k,e}} \hat{a}_\text{k,in} + \sqrt{2\kappa_{k,i}} \hat{a}_\text{k,in,i},\\
\sqrt{2\gamma} \hat{b}^\prime_\text{in} &=& \sqrt{2\gamma} \hat{b}_\text{in} + \sqrt{2\gamma_i} \hat{b}_\text{in,i},\\
\kappa_k &=& \kappa_{k,i} + \kappa_{k,e},\\
\gamma &=& \gamma_i + \gamma_e.
\end{eqnarray}
As the fluctuations in the fields are of primary interest, each Heisenberg operator can be rewritten as a fluctuation term around a steady-state value,
\begin{eqnarray}
 \hat{a}_1(t) &\rightarrow& (\alpha_1 +\hat{a}_1)e^{-i\omega_1 t},\\
 \hat{a}_2(t) &\rightarrow& (\alpha_2 +  \hat{a}_2)e^{-i\omega_2 t},\\
 \hat{b}(t) &\rightarrow& (\beta +  \hat{b})e^{-i\Omega t}. 
\end{eqnarray}
Assuming that the pump beam is driven resonantly with $a_{1}$ ($\omega_L = \omega_1$), the $c$-number steady-state values are equal to $(\alpha_1, \alpha_2, \beta) = (\frac{\sqrt{2 \kappa_{1,e}}}{\kappa} E_{\text{pump}},0,0)$.

For the fluctuation dynamics, with $\Delta \equiv (\omega_{2}-\omega_{1}) - \Omega$, the resulting equations are:

\begin{eqnarray}
 \dot{\hat{a} }_1 &=& - \kappa_1  \hat{a}_1  -  i  h \hat{a}_2 (\hat{b} e^{-i 2 \Omega t}  + \hat{b}^\dagger)   e^{-i\Delta t} -\sqrt{2\kappa} \hat{a}^\prime_\text{1,in}e^{i\omega_1 t} \label{eqn:HL_full_a1dot} \\
\dot{\hat{a} }_2 &=& - \kappa_2 \hat{a}_2  - i h ( \alpha_1 +   \hat{a}_1)  (\hat{b} + \hat{b}^\dagger  e^{+i 2 \Omega t}  )e^{+i\Delta t}   -\sqrt{2\kappa} \hat{a}^\prime_\text{2,in}e^{i \omega_2 t}  \label{eqn:HL_full_a2dot}  \\
\dot{\hat{b}} &=&-\gamma (\hat{b} - \hat{b}^\dagger e^{+i 2 \Omega t} )-i h ( \alpha_1 + \hat a_1 )  \hat a^\dagger_2 e^{+i(\Delta + 2 \Omega)t}
\nonumber \\
&&- i h ( \alpha_1 + \hat a_1 )^\dagger \hat{a}_2 e^{-i\Delta t}   - \sqrt{2\gamma} \hat{b}^\prime_\text{in}e^{i\Omega t} \label{eqn:HL_full_bdot}  
\end{eqnarray}
Ignoring all the mechanically anti-resonant terms for now, and invoking the rotating wave approximation (RWA) valid when $\Delta \ll \Omega$ and $\Omega \gg |\alpha_1| h$, we arrive at the simplified set of fluctuation equations, 
\begin{eqnarray}
 \dot{\hat{a} }_1 &=& - \kappa_1  \hat{a}_1  -  i  h  \hat{a}_2 \hat{b}^\dagger e^{-i\Delta t} -\sqrt{2\kappa} \hat{a}^\prime_\text{1,in} e^{i\omega_1 t}\\
\dot{\hat{a} }_2 &=& - \kappa_2 \hat{a}_2  - i h(\tilde \alpha_1 +   \hat{a}_1) \hat{b}e^{+i\Delta t} -\sqrt{2\kappa} \hat{a}^\prime_\text{2,in} e^{i \omega_2 t} \\
\dot{\hat{b}} &=&  -\gamma \hat{b}  - i h(\tilde \alpha_1 + \hat a_1 )^\dagger  \hat{a}_2  e^{-i\Delta t}   -\sqrt{2\gamma} \hat{b}^\prime_\text{in}e^{i\Omega t} 
\end{eqnarray}

\label{ss:spont_emission}

By ignoring all of the counter-rotating terms proportional to $e^{+i2\Omega t}$, we have also neglected the noise processes alluded to previously due to the $\alpha_{2,-}$ sideband.  Of the mechanically anti-resonant terms which have been dropped, the terms proportional to $\alpha_1$ ($h \alpha_1\hat{b}^\dagger  e^{+i 2 \Omega t}$ in equation~(\ref{eqn:HL_full_a2dot}) and $h  \alpha_1 \hat a^\dagger_2 e^{+i 2 \Omega t}$ in equation~(\ref{eqn:HL_full_bdot})) are the largest and most significant in terms of error in the rotating wave approximation.  These terms correspond to $\hat{a}_1 \hat{a}^\dagger_2 \hat{b}^\dagger$ + h.c. in the Hamiltonian and cause inter-sideband photon scattering between the pump, $\alpha_{1,0}$, and its lower frequency sideband, $\alpha_{2,-}$ as shown in Figure~\ref{fig:doublecav_sidebands}.  This inter-sideband scattering process causes emission and absorption of phonons in the mechanical part of the PPT, thus in principle, even when the extrinsic phonon inputs are in the vacuum state, spontaneous scattering of photons from $\alpha_{1,0}$ to $\alpha_{2,-}$ may populate the mechanical cavity with a phonon. 

This effect was studied in Refs.~\cite{Wilson-Rae2007,marquardt07} in the context of quantum limits to optomechanical cooling.  Similar to that work, a master equation for the phononic mode with the $\omega_1-\Omega$ optical sideband adiabatically eliminated results in an additional \emph{phononic} spontaneous emission term given by,
\begin{equation}
\dot{\rho}_\text{b,spon} =  \frac{G^2}{\kappa}\frac{1}{\left(\frac{2 \Omega}{\kappa}\right)^2 + 1}( 2 \hat{b}^\dagger \rho \hat{b} -  \hat{b}  \hat{b}^\dagger \rho - \rho \hat{b}  \hat{b}^\dagger )
\end{equation}
where $G = h |\tilde \alpha^{\mathrm{ss}}_1|$. The master equation for the mechanics, found by tracing over all bath and optical variables, is of the form $\dot{\rho} = \dot{\rho}_\text{b,i} +\dot{\rho}_\text{b,spon} +\dot{\rho}_\text{b,e}-i/\hbar[H_b,\rho]$, where the terms on the right hand side of the equation are respectively the intrinsic phononic loss, the phonon spontaneous emission, the phonon-waveguide coupling, and the coherent evolution of the system. The first two terms can be lumped together into an effective intrinsic loss,  $\dot{\rho}^\prime_\text{b,i}=\dot{\rho}_\text{b,i} +\dot{\rho}_\text{b,spon}$,
\begin{eqnarray}
\dot{\rho}^\prime_\text{b,i} &=& \gamma_i^\prime (\bar{n}^\prime + 1) ( 2 \hat{b} \rho \hat{b}^\dagger -  \hat{b}^\dagger  \hat{b} \rho - \rho \hat{b}^\dagger  \hat{b}) 
+ \gamma_i^\prime \bar{n}^\prime ( 2 \hat{b}^\dagger \rho \hat{b} -  \hat{b}  \hat{b}^\dagger \rho - \rho \hat{b}  \hat{b}^\dagger ),\label{eqn:effective_louivillian} \\
\gamma_i^\prime &=& \gamma_i (1-n_\text{spon}), \label{eqn:gamma_prime} \\
\bar{n}^\prime &=& \frac{\bar{n} + n_\text{spon}}{1-n_\text{spon}} \label{eqn:n_prime},
\end{eqnarray}
and
\be
n_\text{spon} \equiv \frac{G^2}{\kappa \gamma_i} \frac{1}{\left(\frac{2 \Omega}{\kappa}\right)^2 + 1}.
\ee
Hence, by assuming that the intrinsic loss phonon bath is at a modified temperature with occupation number $\bar{n}^\prime$, and changing the intrinsic phonon loss rate to $\gamma^\prime_i$, the spontaneous emission and intrinsic loss noise are lumped into one effective thermal noise Liouvillian for the mechanics.  Note that it is possible in this model to have $\gamma^\prime_i$ negative when $n_\text{spon}>1$; however, the motional decoherence rate is always positive and given by $\gamma_i^\prime \bar{n}^\prime = \gamma_i (\bar{n} + n_\text{spon})$.

In Section~\ref{ss:scattering_matrix} we will see that the optimal state transfer efficiency is given by $G^2 = \gamma \kappa$ \footnote{The $\gamma$ in this case is in principle $\gamma_e + \gamma^\prime_i$ as opposed to $\gamma_e + \gamma_i$, and the equations must therefore be solved self-consistently. This is discussed in section~\ref{ss:spont_mod}.} in which case\begin{equation}
n_\text{spon} \approx  \frac{\gamma}{\gamma_i} \left(\frac{\kappa}{2 \Omega}\right)^2,
\end{equation}
\noindent in the sideband resolved limit.  In the case where $g \ne 0$ (recall this is the self-coupling radiation pressure term), photons scattering from the pump into the lower frequency sideband ($\omega_1 - \Omega$) can scatter into the $a_{1}$ cavity mode which is only detuned by $\Omega$.  This is to be compared with the $g=0$ case considered thus far in which the detuning is $2\Omega$ for scattering into the $a_{2}$ mode.  As such, for the $g \ne 0$ case there will be roughly four times the spontaneous emission noise, with $n^{g\ne0}_\text{spon} \approx  {\gamma}/{\gamma_i} ({\kappa}/{\Omega})^2$.

\begin{figure}[htbp]
\begin{center}
\scalebox{1}{\includegraphics{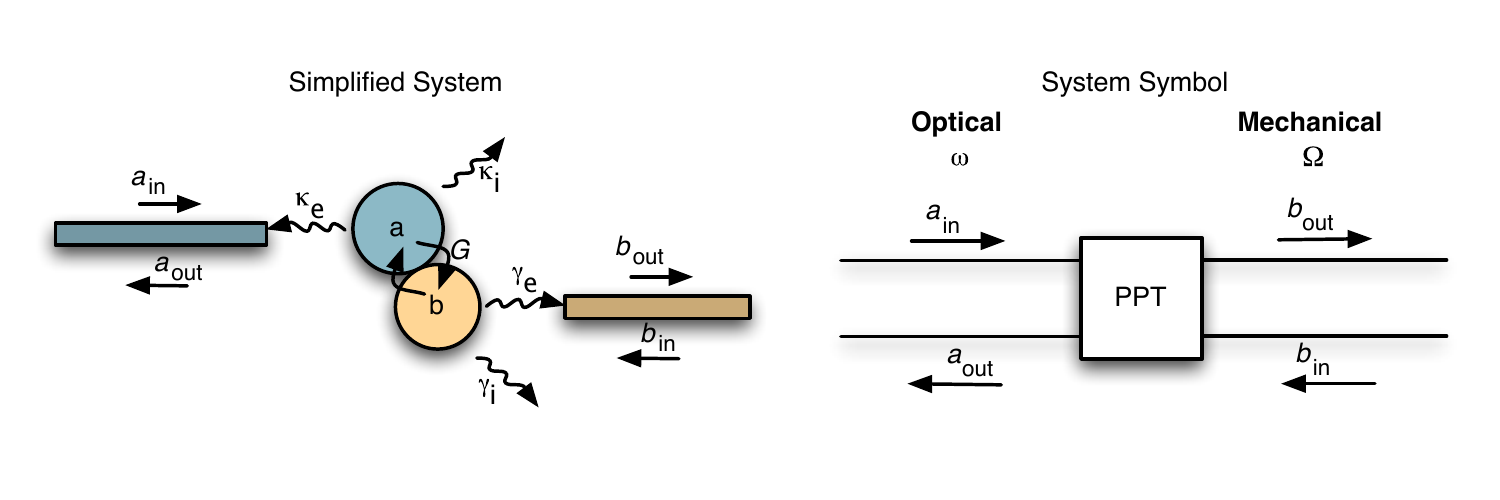}}
\caption{Simplified PPT system diagram and symbol. The coupling rate $G$ is proportional to $h$ and $\sqrt{n_1}$, where $n_1$ is the number of photons in the pump mode. }
\label{fig:simplified_system_diagram}
\end{center}
\end{figure}

A final simplification can be made by neglecting the fluctuations in the strong optical pump of cavity mode $a_{1}$.  Considering that the fluctuations in the variables are all of the same order, and that $ \hat{a}_1$ always appears as $\alpha_1 +   \hat{a}_1$ in the equations of motion for $\hat{b}$ and $\hat{a}_2$, we can ignore the dynamics of the pump fluctuations in the case where $|\alpha_1| \gg \avg{\op{a}{1}},\avg{\op{a}{2}}$ and $\avg{\op{b}{}}$. This is the undepleted pump approximation. Adiabatically removing the pump from the dynamics of the system yields a pump-enhanced optomechanical coupling $G = h |\alpha_1|$ between optical cavity mode $a_{2}$ and the mechanical resonance $b$.  Dropping the subscript from the cavity mode $a_{2}$ and moving to a rotating reference frame results in the new effective Hamiltonian~\cite{Zhang2003},
 \begin{equation}
\hat{H}_{\mathrm{eff}} = -\Delta \hat{b}^\dagger \hat{b}+
G \hat{a}^\dagger \hat{b} + G^\ast \hat{a} \hat{b}^\dagger.
\label{eqn:H_eff_start}
\end{equation}
\noindent The system diagram and symbol corresponding to this simplified model of the PPT are shown in Figure~\ref{fig:simplified_system_diagram}.

\subsection{Scattering Matrix Formulation of the Phonon-Photon Translator}\label{ss:scattering_matrix}

To understand the properties of the PPT as a waveguide adapter, we begin with a study of its scattering matrix. Starting from the effective Hamiltonian given in equation~(\ref{eqn:H_eff_start}),  the Heisenberg-Langevin equations of motion for the Hamiltonian~(\ref{eqn:H_eff_start}) are written under a Markov approximation in the frequency domain,
\begin{eqnarray}
 - i \omega \atl &=& -\kappa \atl - i G \btl - \sqrt{2 \kappa_e} \atlss{in} - \sqrt{2 \kappa_i} \atlss{in,i}, \label{eqn:freqdom1}\\
 - i \omega \btl &=& -(\gamma - i \D) \btl - i G^\ast \atl  - \sqrt{2 \gamma_e} \btlss{in} - \sqrt{2 \gamma_i} \btlss{in,i}\label{eqn:freqdom2}.
\end{eqnarray}
The intrinsic noise terms $\atlss{in,i}$ and $\btlss{in,i}$ are the initial-state boson annihilation operators for the baths, while the extrinsic terms $\atlss{in}$ and $\btlss{in}$ are annihilation operators for the optical and mechanical guided modes for each respective waveguide.  Since the effective Hamiltonian (\ref{eqn:H_eff_start}) has been used to derive equations (\ref{eqn:freqdom1}-\ref{eqn:freqdom2}), thus neglecting the counter-rotating terms present in the full system dynamics, the effects of phonon spontaneous emission noise is included separately.  Following the discussion in Section~\ref{ss:spont_emission}, the effective Liouvillian~(\ref{eqn:effective_louivillian}) corresponds to replacing $\gamma_i$ with $\gamma_i^\prime$ in (\ref{eqn:freqdom2}) and using a Langevin force $\btlss{in,i}$ satisfying the relations,
\begin{eqnarray}
\avg{\btlssd{in,i}\btlssp{in,i}} &=&\bar{n}^\prime \delta(\omega-\omega^\prime), \\
\avg{\btlssp{in,i}\btlssd{in,i}} &=& (\bar{n}^\prime+1) \delta(\omega-\omega^\prime),
\end{eqnarray}
where $\gamma_i^\prime$ and $\bar{n}^\prime$ are given in equations (\ref{eqn:gamma_prime}-\ref{eqn:n_prime}). The intrinsic optical noise correlations are only due to vacuum fluctuations and given by $\avg{\atlssp{in,i}\atlssd{in,i}} = \delta(\omega-\omega^\prime)$. 

 To ensure efficient translation, competing requirements of matching and strong coupling between waveguide and resonator must be satisfied. This is similar to the problem of designing integrated optical filters using resonators and waveguides~\cite{Manolatou1999,Barclay2003}. From the above equations and the input-output boundary condition ~\cite{Gardiner1993,GardinerZoller}, we arrive at the matrix equation,
\begin{eqnarray}
\twovec{\atlss{out}}{\btlss{out}} &=&  \mathcal{S} \twovec{\atlss{in}}{\btlss{in}} + \mathcal{N}\twovec{\atlss{in,i}}{\btlss{in,i}},
\end{eqnarray}
with scattering and noise matrices
\begin{equation}
\mathcal{S} = \twotwomat{s_{11}(\omega)}{s_{12}(\omega)}{s_{21}(\omega)}{s_{22}(\omega)}\qquad\mbox{and}\qquad
\mathcal{N} = \twotwomat{n_{11}(\omega)}{n_{12}(\omega)}{n_{21}(\omega)}{n_{22}(\omega)}.
\end{equation}
The elements of the scattering matrix $\mathcal{S}$ are
\begin{eqnarray}
 s_{11}(\omega) &=& 1 - \frac{ 2 \kappa_e (\gamma -i(\D + \omega)) }{|G|^2 + (\gamma - i(\omega+\D))(\kappa - i \omega)},\\
 s_{12}(\omega) &=& \frac{ 2 i G^\ast \sqrt{\gamma_e \kappa_e}}{ |G|^2 + (\gamma - i(\omega+\D))(\kappa - i \omega)},\\
 s_{21}(\omega) &=& s_{12}^\ast (\omega),\\
 s_{22}(\omega) &=& 1 - \frac{ 2 \gamma_e (\kappa -i\omega) }{|G|^2 + (\gamma - i(\omega+\D))(\kappa - i \omega)}.
\end{eqnarray}
Similar expressions are also found for the noise scattering matrix elements $n_{ij}(\omega)$, with their extrema reported below.

\begin{figure}[htbp]
\begin{center}
\scalebox{1}{\includegraphics{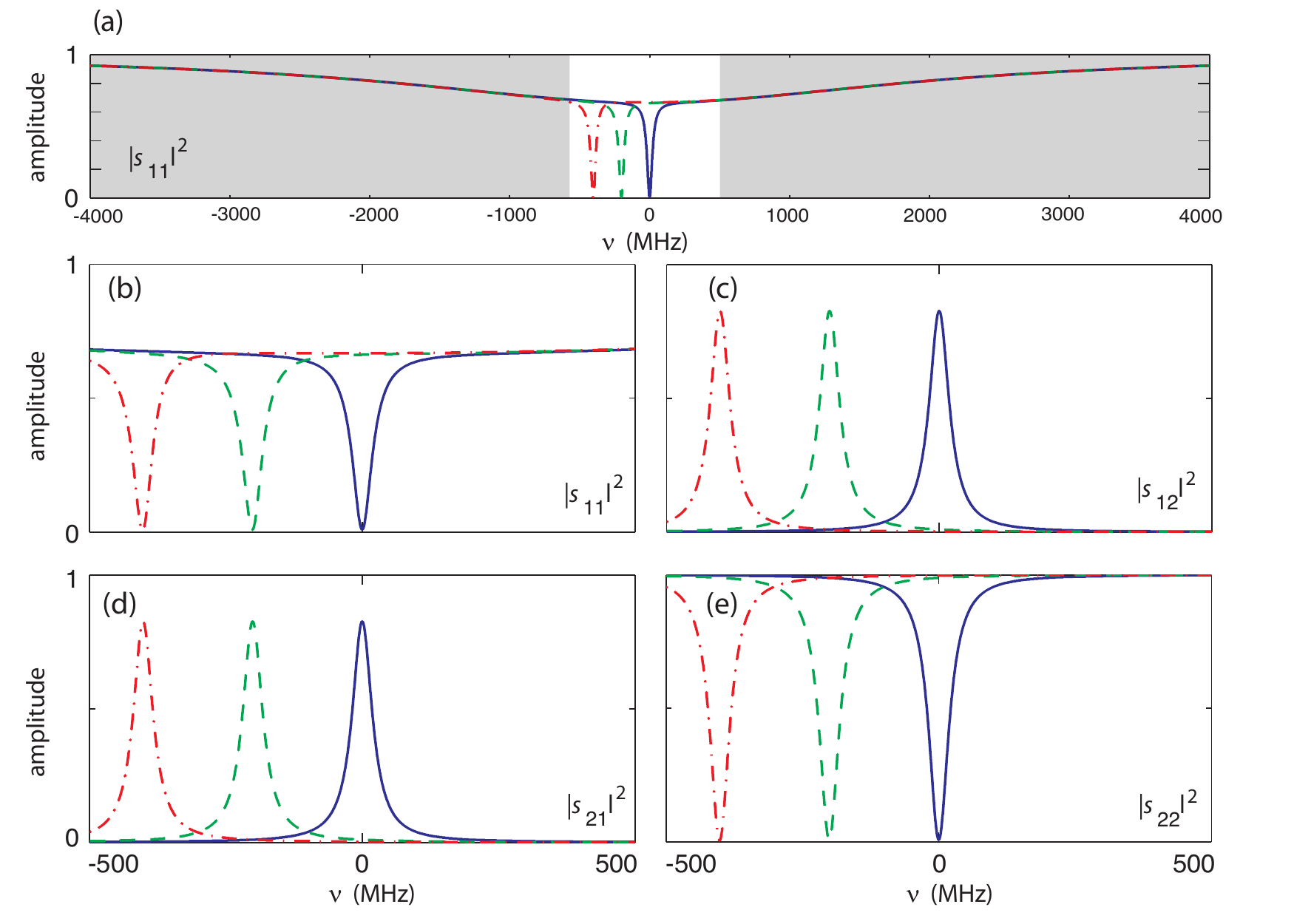}}
\caption{Phonon-photon scattering matrix amplitudes for  $(\gamma_e,\gamma_i,\kappa_e,\kappa_i,G)=2\pi\times(10,1,2000,200,155.6) ~\text{MHz}$. (a) Plot of the frequency dependence of the optical reflection $s_{11}$. The broad over-coupled optical line is visible, along with the PPT feature near the center in the unshaded region. This unshaded region is shown in more detail in plots (b), (c), (d),  and (e), showing the frequency dependence of the scattering matrix elements $s_{11}$, $s_{12}$, $s_{21}$ and $s_{22}$, respectively. In each plot, the curves (\textcolor{blue}{$-$}), (\textcolor{green}{$--$}) and (\textcolor{red}{$\cdot -$}) represent detunings of $\Delta = 0, 2\pi\times 200$ and $2\pi\times 400$ MHz respectively. }
\label{fig:scattering_matrix}
\end{center}
\end{figure}

In order to obtain efficient conversion the cavities must be over-coupled to their respective waveguides, ensuring that the phonon (photon) has a higher chance of leaking into the waveguide continuum modes than escaping into other loss channels. In this regime, $\kappa \approx \kappa_e$ and $\gamma \approx \gamma_e$.  In the weak coupling regime, $G < \kappa$, the response of the system exhibits a maximum for $s_{12}$ and $s_{21}$ at $\omega=0$ and a minimum at the same point for $s_{11}$ and $s_{22}$. In fact, with realistic system parameters, only the weak-coupling regime leads to efficient translation. In strong-coupling ($G > \kappa$ and $\kappa \gg \gamma$) the photon is converted to a phonon at rate $G$, and then back to a photon before it has a chance to leave through the much slower phononic loss channel at rate $\gamma$, causing there to be significant reflections and reduced conversion efficiency. To find the optimal value of $G$ we consider the extrema given by
\begin{eqnarray} 
 |s_{11}|_\text{min} &=& \left| \frac{G^2 + \gamma \kappa_i - \gamma \kappa_e}{G^2 + \gamma \kappa_i + \gamma \kappa_e}\right|,\\
 |s_{12}|_\text{max} &=& \left| \frac{2 G \sqrt{\gamma_e \kappa_e}}{G^2 + \gamma \kappa}\right|,\\
 |s_{22}|_\text{min} &=& \left| \frac{G^2 + \kappa \gamma_i - \kappa \gamma_e}{G^2 + \kappa \gamma_i + \kappa \gamma_e}\right|.
\end{eqnarray}
In the over-coupled approximation and in the case where $\kappa_i=\gamma_i=0$, it is easy to see that the full translation condition $|s_{12}|_\text{max} = 1$ is achievable by setting $G$ equal to 
\be
G^\text{o} = \sqrt{\gamma \kappa}\label{eqn:matching_condition}.
\ee
This result has a simple interpretation as a matching requirement. The photonic loss channel viewed from the phononic mode has a loss rate of $\frac{G^2}{\kappa}$. Matching this to the purely mechanical loss rate of the same phononic mode, $\gamma$, one arrives at $G^\text{o} = \sqrt{\gamma \kappa}$. The same argument can be used for the photonic mode, giving the same result.  Under this matched condition, the linewidth of the translation peak in $|s_{12}|^2$ is simply
\begin{equation}
 \gamma_\text{transfer} = \frac{4 |G^\text{o}|^2}{\kappa} = 2\gamma. \label{eqn:gamma_transfer}
\end{equation}

With intrinsic losses taken into account, either $|s_{11}|$ or $|s_{22}|$ (but not both) can be made exactly 0 by setting $G^2 = \gamma(\kappa_e - \kappa_i)$ or $G^2 = (\gamma_e - \gamma_i)\kappa$, respectively.   The optimal state transfer condition, however, still occurs for $G^\text{o} = \sqrt{\gamma \kappa}$.  The extremal values ($\omega = 0$) of the scattering matrix are in this case are,
\begin{eqnarray}
 |s_{11}|^{\text{optimal}}_{\omega=0} &=& \frac{\kappa_i}{\kappa_e + \kappa_i}, \label{eqn:s11}\\
 |s_{12}|^{\text{optimal}}_{\omega=0}&=& \sqrt{\frac{\gamma_e \kappa_e}{\gamma \kappa}},\\
 |s_{22}|^{\text{optimal}}_{\omega=0} &=& \frac{\gamma_i}{\gamma_e + \gamma_i},
\end{eqnarray}
\noindent with corresponding noise matrix elements of,
\begin{eqnarray}
 |n_{11}|^{\text{optimal}}_{\omega=0} &=& \frac{\sqrt{\kappa_i \kappa_e}}{\kappa_e + \kappa_i},\label{n_mat_1}\\
 |n_{12}|^{\text{optimal}}_{\omega=0}&=& \sqrt{\frac{\gamma_i \kappa_e}{\gamma \kappa}},\label{n_mat_2}\\
 |n_{21}|^{\text{optimal}}_{\omega=0}&=& \sqrt{\frac{\gamma_e \kappa_i}{\gamma \kappa}},\label{n_mat_3}\\
 |n_{22}|^{\text{optimal}}_{\omega=0}&=& \frac{\sqrt{\gamma_i \gamma_e}}{\gamma_e + \gamma_i}.\label{n_mat_4}
\end{eqnarray}

For a set of parameters typical of an optomechanical crystal system, the magnitudes of the scattering matrix elements versus frequency are plotted in Figure~\ref{fig:scattering_matrix}.  In these plots we have assumed resonant optical pumping of the $a_{1}$ cavity mode and considered several different detuning values $\Delta$.  The normalized optical reflection spectrum ($|s_{11}|^2$) is shown in Figure~\ref{fig:scattering_matrix}(a), in which the broad optical cavity resonance can be seen along with a deeper, narrowband resonance that tunes with $\Delta$. This narrowband resonance is highlighted in Figure~\ref{fig:scattering_matrix}(b), showing that the optical reflection is nearly completely eliminated on resonance.  Photons on resonance, instead of being reflected, are being converted into outgoing phonons as can be seen in resonance peak of $|s_{21}|^2$ shown in Figure~\ref{fig:scattering_matrix}(d). A similar reflection dip and transmission peak is visible for the phononic reflection ($s_{22}$) and phonon to photon translation ($s_{12}$) curves. It is also notable from Figure~\ref{fig:scattering_matrix} that for small detunings $\Delta$ of the system ($\Delta < \kappa, \Omega$) the resonant scattering matrix elements are only weakly affected and the translation process maintains its efficiency.


\subsubsection{Modifications to Matching Condition due to Counter-Rotating Terms}\label{ss:spont_mod}
As a consequence of the counter-rotating terms treated in Section~\ref{ss:spont_emission}, $\gamma$ is weakly dependent on $G$.  In particular, by making the substitution $\gamma_i \rightarrow \gamma_i^\prime$, where $\gamma_i^\prime$ is given by equation~(\ref{eqn:gamma_prime}), the equation for the optimal $G$ becomes $\frac{|G^\text{o}|^2 }{\kappa}=\gamma_e + \gamma_i^\prime$, where $\gamma_i^\prime$ is itself dependent on $G$. Algebraic manipulations give us the desired value of $G$, 
\begin{equation}
|G^\text{o}|^2 = \left(\gamma \kappa \right)\frac{\left(1+\left(\frac{2\Omega}{\kappa}\right)^2\right)}{\left(2+\left(\frac{2\Omega}{\kappa}\right)^2\right)}, \label{eqn:Go_mod}
\end{equation}
simplifying in the sideband-resolved limit to $G^\text{o} = \sqrt{\gamma \kappa}\left(1 - \left(\frac{\kappa}{2\Omega}\right)^2\right).$  The values for the scattering matrix elements given in equations (\ref{eqn:s11}-\ref{n_mat_4}) are also suitably modified by the substitutions $\gamma_i \rightarrow \gamma_i^\prime$ and $\gamma \rightarrow \gamma_e + \gamma_i^\prime$.


\subsubsection{Phononic Waveguide Losses}\label{app:waveguide_loss}

The issue of waveguide loss is one that is normally ignored in quantum optical systems where low-loss fiber or free-space links are readily available. On the phononic side of the systems studied here, the length of the waveguide and its intrinsic losses may be large enough for the waveguide attenuation factor to become important. Because of the negative effect of attenuation in applications involving optomechanical delay lines, and quantum state transfer, it is useful to model this loss and see how the scattering matrix elements are altered.

\begin{figure}[htbp]
\begin{center}
\scalebox{1}{\includegraphics{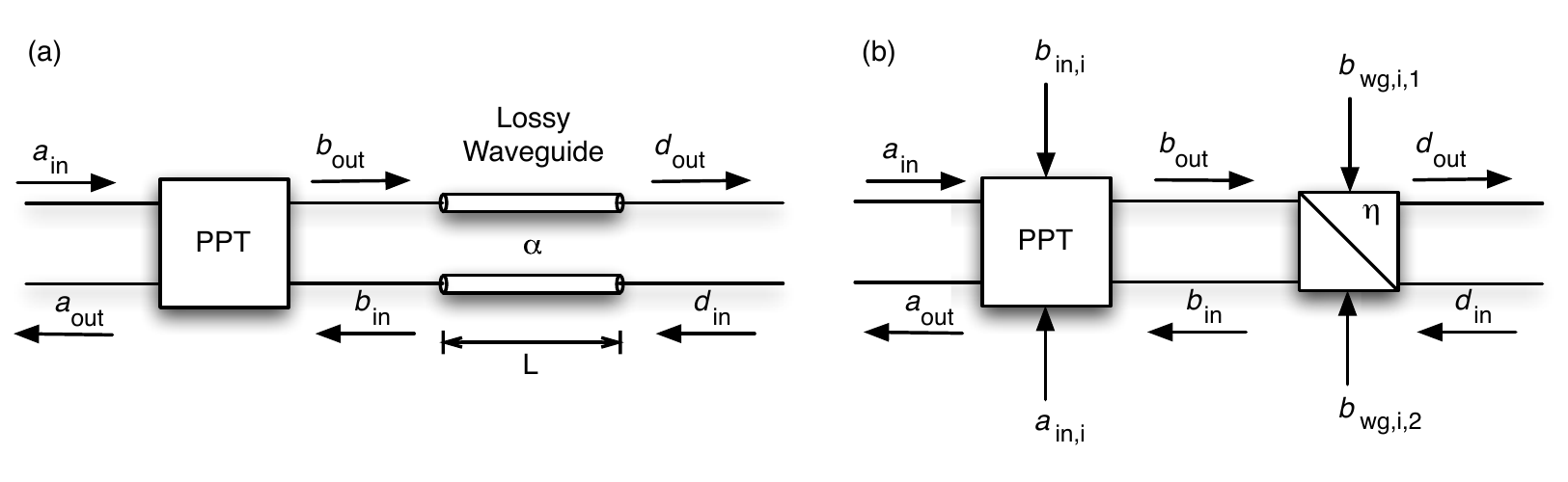}}
\caption{(a) System diagram of a PPT connected to a lossy waveguide of length $L$ with attenuation factor $\alpha$. (b) An equivalent system diagram in which the waveguide is replaced by a beam splitter with a reflectivity $\eta$. }
\label{fig:waveguide-loss}
\end{center}
\end{figure}

A model of the system shown in Figure~\ref{fig:waveguide-loss}(b), where the lossy waveguide is replaced by a single beam splitter. This is accomplished by first modeling the lossy waveguide as a large number $N$ of cascaded beam splitters, each reflecting $2\alpha \Delta z$ away from the main beam, with $L = \Delta z N$. This serial array of beam splitters can be combined into one, with reflectivity $\eta = e^{-2\alpha L}$ as $N\rightarrow \infty$.
The relation for $\hat{d}_\text{out}$ can be found by starting from the approximate scattering matrix relation for $\hat{b}_\text{out}$,
\begin{eqnarray}
\hat{b}_\text{out}(t) &=&  s_{21}\hat{a}_\text{in}(t) + s_{22} \hat{b}_\text{in}(t)  + n_{12} \hat{a}_\text{in,i}(t)+ n_{22}\hat{b}_\text{in,i}(t),
\end{eqnarray}
and using the beam splitter relations,
\begin{eqnarray}
\hat{d}_\text{out}(t) &=& \sqrt{\eta} ~b_\text{out} + \sqrt{1-\eta}~b_\text{wg,i,1}\\
\hat{b}_\text{in}(t) &=& \sqrt{\eta} ~d_\text{in} + \sqrt{1-\eta}~b_\text{wg,i,2}.
\end{eqnarray}
Assuming that the phonon cavity intrinsic loss bath is at the same temperature and uncorrelated with the phonon waveguide intrinsic loss baths, then one finds for translation through the lossy phonon waveguide,
\begin{eqnarray}
\hat{d}_\text{out}(t) &=&  s^\prime_{21}\hat{a}_\text{in}(t) + s^\prime_{22} \hat{d}_\text{in}(t) + n^\prime_{12} \hat{a}_\text{in,i}(t)+ n^\prime_{22}\hat{b}_\text{in,i}(t),
\end{eqnarray} 
where 
\begin{eqnarray}
s^\prime_{21} &=&\sqrt{\eta}~s_{21},\\
s^\prime_{22} &=&\eta~s_{22},\\
n^\prime_{12} &=&\sqrt{\eta}~n_{12},\\
n^\prime_{22} &=&\sqrt{(1-\eta)\eta |s_{22}|^2 + \eta|n_{22}|^2 +1-\eta}.
\end{eqnarray}
The value of $s_{21}$ is simply reduced by a factor $\sqrt{\eta}$ due to the lossy waveguide.  For propagation lengths short relative to the attenuation length of the lossy waveguide, $\alpha L \ll 1$,  and the reduction in translation $\sqrt{\eta} \approx 1-\alpha L$ is small.  The added noise due to the waveguide attenuation is contained in $n^\prime_{22}$, and is also seen to be small for $\alpha L \ll 1$.

\subsection{Effects of Thermal and Quantum Noise}\label{ss:noise}

In general, the type of noise which is relevant to the PPT depends on the conditions in which it is used. For example, when used as a bridge between RF-microwave photonics and optics for classical applications at room temperature, the thermal noise affecting the RF signal will be at a level which makes the quantum noise induced by spontaneous pump scattering irrelevant. On the other hand, when the system is used at cryogenic temperatures for connecting a superconducting circuit QED system to an optical system as described below, the quantum noise of the translation process itself becomes dominant. In what follows we analyze both sources of noise.

\subsubsection{Thermal Noise on the Optical Side}

The noise power on each of the output waveguide channels can be found by using the scattering matrix formulation described in Section~\ref{ss:scattering_matrix}. Here we evaluate the effects of thermal noise classically. When the system is used an optical drop filter, in which an optical beam is sent in and the optical reflection is measured, we find:
\begin{eqnarray}
a_\text{out}(t) &=& s_{11}(t) \convolution a_\text{in}(t) +  s_{12}(t) \convolution b_\text{in}(t) \nonumber\\
&& + n_{11}(t) \convolution a_\text{in,i} (t)+n_{12}(t) \convolution b_\text{in,i}(t)
\end{eqnarray}
where $\convolution$ represents convolution. Assuming that there are no cross-correlations between the various input noise terms, we find the classical spectral density of the noise to be:
\begin{eqnarray}
S_\text{out}(\omega) =  |s_{12}(\omega)|^2 \bar{n} +  |n_{12}(\omega)|^2 (\bar{n} + n_\text{spon}).\label{eqn:Sout1}
\end{eqnarray}
For a system with mechanical frequency less than $10$~GHz, at room temperature ($T=300~\text{K}$) the corresponding spontaneous emission noise is much less than the thermal noise, $n_\text{spon} \ll \bar{n} \approx 10^3$, and $n_\text{spon}$ can be ignored.
 
Ignoring the quantum noise term for the moment, and evaluating equation (\ref{eqn:Sout1}) at $\omega = 0$ after substituting in equations (\ref{n_mat_1}-\ref{n_mat_4}), the total thermal noise power on the reflected optical signal is found to be, 
\begin{equation}
P^N_{\text{o,out}} = \hbar \omega_2 \bar{n} \frac{\kappa_e}{\kappa} \pi B, \label{eqn:P_nbar}
\end{equation}
where $B=2\gamma$ is the bandwidth of the PPT.  In the classical regime, $\bar{n} = kT_\text{bath} / (\hbar \Omega)$, and this equation reduces to $ P^N_{\text{o,out}}  = 2\pi \gamma  kT_\text{bath}({\omega_2}/{\Omega})({\kappa_e}/{\kappa}).$ 
This result has a simple interpretation. The noise power $\pi\gamma kT_\text{bath}$ is the standard thermal noise input on the phononic side of the PPT. The ratio $\omega_2/\Omega$ is a translation factor, arising from the fact that quantum-limited conversion of phonons to photons causes an increase in energy by the factor of the ratio of their frequencies. Finally,  the factor $\kappa_e/\kappa$ is the extraction efficiency of photons from the optical mode to the waveguide.
Alternatively, one may define an equivalent optical temperature by setting $\pi \kappa_e k T_\text{o,eff} = P^N_{\text{o,out}}$, yielding $T_\text{o,eff} =  ({2 Q_o}/{Q_m})T_\text{bath}$, where the $Q$'s represent the loaded quality factors of the optical and mechanical resonators. This last expression must be interpreted carefully, only in terms of a power equivalence, as spectrally the noise on the optical side of the PPT has a bandwidth of $2\gamma$, while thermal noise radiating from the cavity would have a bandwidth of $\kappa$.

\subsubsection{Phonon Spontaneous Emission Noise}

To calculate the effective increase in noise brought about by the spontaneous pump scattering and phonon emission process, we start from equation~(\ref{eqn:Sout1}) and use the spontaneous emission contribution $n_\text{spon}$ to find,
\begin{equation}
P^N_{\text{o,out}}  = \left(\hbar \omega_2\right) \left(\bar{n}+  \frac{1}{\left(\frac{2 \Omega}{\kappa}\right)^2 + 1}\right)\left(\frac{\kappa_e}{\kappa}\right) \left(2\pi \gamma\right). 
\end{equation}
The second term in the brackets is due to the spontaneous emission of phonons by the optical pump beam, and is added to the thermal noise exiting the optical side of the PPT.  Put in terms of an effective contribution to the thermal \emph{photon} occupation number, the spontaneous pump scattering effectively adds, 
\begin{eqnarray}
n_\text{o,spon} = \frac{2 \gamma}{\kappa} \left(\frac{\kappa}{2 \Omega}\right)^2,
\end{eqnarray}
\noindent thermal photons to the cavity.  This equivalence is only in terms of total noise power emitted, as spectrally the noise is emitted over the $2\gamma$ bandwidth of the PPT resonance, not the entire optical cavity resonance as discussed above.  

\section{Proposed On-Chip Implementation} \label{sec:implementation}

Up to this point in the analysis of the PPT, the discussion has been kept as general as possible. Such a system is, however, interesting only insofar as it is realizable, and we attempt here to establish the practicality of a PPT. Building upon recent experimental~\cite{Eichenfield2009} and theoretical work~\cite{Eichenfield2009a,Safavi-Naeini2010}, we provide the design of an optomechanical crystal formed in a silicon microchip that can realize a PPT system with high phonon-photon translation efficiency.

As previously mentioned, optomechanical crystals~\cite{Eichenfield2009} are engineered structures in which phonons and photons may be independently routed and their interactions controlled. In order to create a suitable OMC structure for the implementation of a PPT device one looks to a crystal lattice providing simultaneous phononic and photonic bandgaps for the guiding and co-localization of phonons and photons~\cite{Maldovan2006}.  We have recently proposed~\cite{Safavi-Naeini2010} such an OMC system, formed from a silicon-on-insulator wafer and consisting of a patterned thin membrane of silicon.  The proposed ``snowflake'' crystal lattice supports a phononic bandgap in the $5$-$10$~GHz mechanical frequency band and a photonic bandgap in the $1500$~nm optical wavelength band.  This quasi-2D crystal structure was also shown to support low-loss photonic and phononic waveguides, optical resonances with radiation-limited $Q > 10^6$ co-localized with mechanical resonances of frequency $\Omega \approx 2\pi\times 10~\text{GHz}$, and a single quanta optomechanical interaction rate of $g \approx 2\pi \times 300~\text{kHz}$. 

In this section we design an example OMC implementation of a PPT in the silicon snowflake crystal. We limit ourselves here to a two-dimensional (2D) crystal involving only 2D Maxwell's equations for transverse-electric (TE) polarized optical waves and in-plane elastic deformations of an infinitely thick slab.  This simplifies the analysis and avoids some of the technical challenges related to achieving high optical $Q$s in a quasi-2D thin film structure, challenges which have already been studied and met elsewhere~\cite{Song2005a,Kuramochi2006,Safavi-Naeini2010}.

\subsection{Single and Double Cavity Systems}

\begin{figure}[htbp]
\begin{center}
\includegraphics{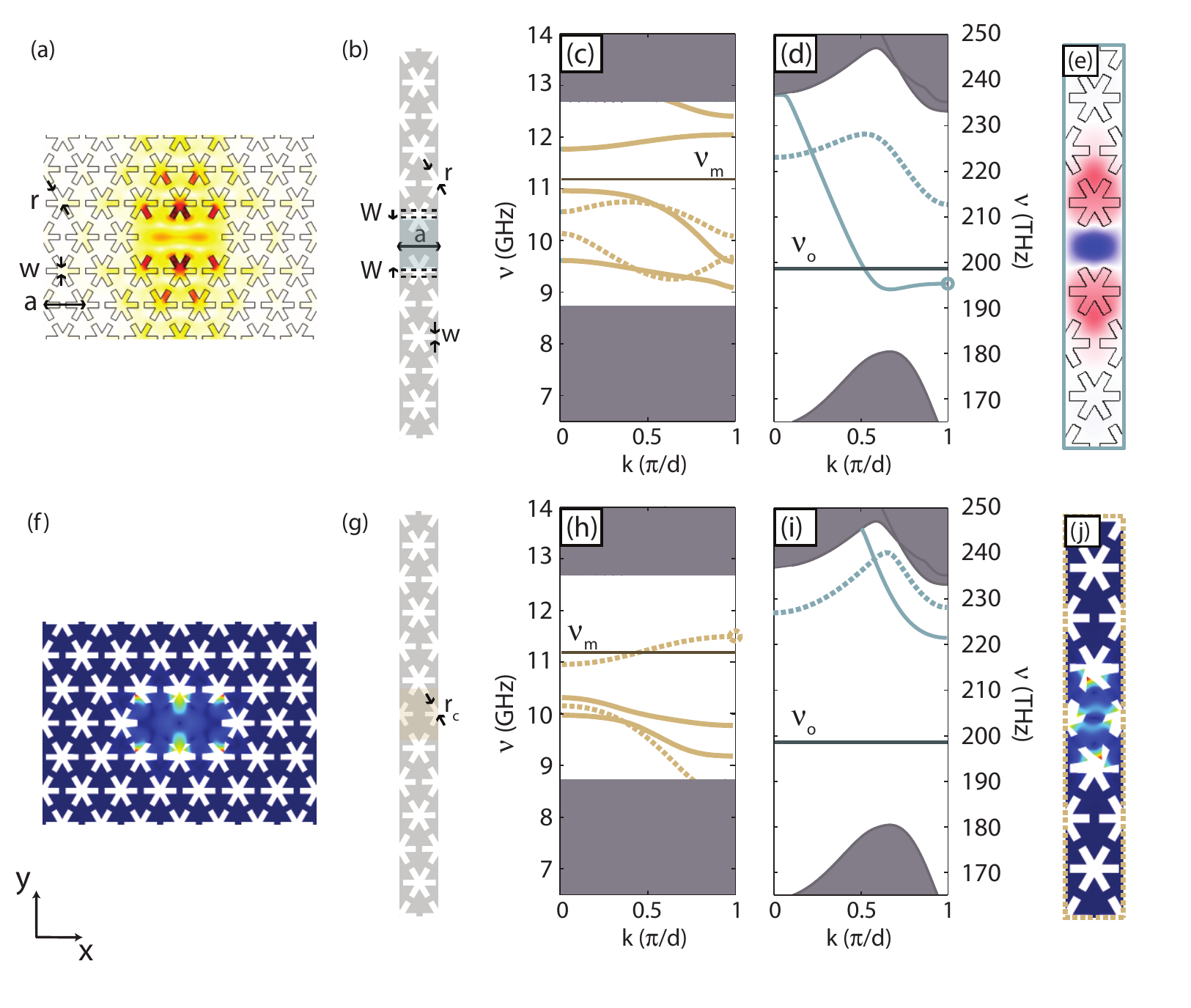}
\end{center}
\caption{(a) The electric field magnitude $|\m E|$ of the L2 cavity photonic resonance at $\nu_o = 199~\text{THz}$. (b-e) The line defect optical waveguide structure, described in the text, which couples to the L2 cavity photonic resonance.  The numerically simulated (c) acoustic bandstructure and (d) optical bandstructure of the line defect optical waveguide with $W=0.135a$. (e) Plot of the out-of-plane component of the magnetic field, $H_z$, of the guided optical mode at $X$-point of the bandstructure. (f) The displacement field $|\m Q|$ of the L2 cavity phononic resonance at $\nu_m = 11.2~\text{GHz}$.  (g-j) The line defect acoustic waveguide structure, described in the text, which couples to the L2 cavity phononic resonance.  The numerically simulated (h) acoustic bandstructure and (i) photonic bandstructure of the line defect acoustic waveguide with $r_c = 0.82 r$.  (j) The magnitude of the mechanical displacement field, $|\m Q|$, of the guided acoustic mode at the $X$-point of the bandstructure.  Calculations of the acoustic waveguide bandstructure are done using an FEM model~\cite{COMSOL2009}, while for the optical waveguide simulations, a plane-wave-expansion method was utilized~\cite{Johnson2001:mpb}.
\label{fig:cavities_and_waveguides}}
\end{figure}

The snowflake crystal consists of a hexagonal lattice of snowflake-shaped holes patterned into silicon as shown in Figure~\ref{fig:cavities_and_waveguides}.  The snowflake lattice used here is characterized by a lattice constant $a=400~\text{nm}$, snowflake radius $r = 168~\text{nm}$, and width $w = 60~\text{nm}$.  It possesses a full phononic bandgap from $8.6$ to $12.6$ GHz and a photonic pseudo-bandgap for TE optical waves from $180$ to $230$ THz.  Defects in the crystal (features breaking the discrete translational symmetry of the underlying lattice) can support resonances with frequencies within the optical and mechanical bandgaps, leading to highly localized, strongly interacting resonances~\cite{Safavi-Naeini2010}.  As an example, by filling an adjacent pair of snowflake-shaped holes in the crystal, a so-called ``L2'' cavity is formed which supports a localized photonic resonance at a frequency $\nu_o = 199~\text{THz}$ and a phononic resonance at a frequency $\nu_m = 11.2~\text{GHz}$.  The defect cavity structure and finite-element-method (FEM)~\cite{COMSOL2009} simulated field profiles of these co-localized modes are shown in Figures~\ref{fig:cavities_and_waveguides}(a) and ~\ref{fig:cavities_and_waveguides}(f). 

The L2 cavity forms the basis of a more complex double-optical-mode cavity structure with the desired symmetry properties for efficient phonon-to-photon translation.  By placing two separate L2 cavities close to one another, at sufficiently small separations, even and odd optical and mechanical super-modes form with split mechanical and optical resonant frequencies. We choose to displace the cavities from each other in the $y$ direction, as shown in full system diagram of Figure~\ref{fig:full_system_only}. These super-modes of the coupled cavities are characterized with respect to their vector symmetry about the mirror symmetry transformation $\sigma_y  (x,y) = (x,-y)$. We denote these super-modes $\m E_\pm$ and $\m Q_\pm$, where `$+$' denotes symmetric and `$-$' denotes anti-symmetric symmetry with respect to $\sigma_y$. The supermodes can be written approximately as,
\be
\m E_\pm = \frac{\m E_a \pm \m E_b}{\sqrt{2}}\qquad\mbox{and}\qquad\m Q_\pm = \frac{\m Q_a \pm \m Q_b}{\sqrt{2}},
\ee
\noindent where the subscripts $a,b$ label the individual cavity fields.  The cavity separation (14 rows) shown in Figure~\ref{fig:full_system_only} is chosen such that the optical supermode frequency splitting is very nearly identical to the mechanical mode frequency of $\nu_m = 11.2~\text{GHz}$, as ascertained by FEM simulations.  We focus here only on the mechanical mode of odd vector symmetry, $\m Q_-$, since as we will show below, this is the mode which cross-couples the two optical super-modes to each other.

\subsection{Optomechanical Coupling Rates}

Optomechanical coupling (or acousto-optic scattering) arises from the coupling of optical cavity modes under deformations in the geometric structure.  In the canonical form of radiation pressure, a mechanical deformation in the cavity induces a shift in the resonance frequency of a given cavity mode.  The coefficient describing the cavity mode dispersion with mechanical displacement also quantifies the strength of the radiation pressure force that photons in the cavity mode exert back on the mechanical structure.  More generally, mechanical deformations may couple one optical cavity mode to another.  

The self-coupling and inter-modal couplings caused by a mechanical deformation are modeled by the position dependent interaction rates ${g}_k(\hat{x})$ and ${h}(\hat{x})$, respectively, in the Hamiltonian of eqn.~(\ref{eqn:full_total_Hamiltonian}).  Both types of deformation-dependent optomechanical couplings may be calculated to first order using a variant of the Feynman-Hellmann perturbation theory, the Johnson perturbation theory~\cite{Johnson2002}, which takes into account moving boundaries in electromagnetic cavities and has been used successfully in the past to model optomechanical crystal cavities~\cite{Eichenfield2009,Eichenfield2009a}. The Hamiltonian, given to first-order by,
\bea
\op{H}{} &=&  \hbar \sum_{i} \omega_{i} \opdagger{a}{i}\op{a}{i} + \hbar \Omega \opdagger{b}{}\op{b}{}+ \frac{\hbar}{2} \sum_{i,j} g_{i,j} (\opdagger{b}{} + \op{b}{}) \opdagger{a}{i} \op{a}{j},
\eea
is then a generalization of that shown previously in equation (\ref{eqn:total_Hamiltonian}), with
\bea
g_{i,j} &= &\frac{\omega_{i,j}}{2} \sqrt{\frac{\hbar}{2\omega_m t}}\frac{ \int dl \left(\m{Q} \cdot \m{n} \right) \left(\Delta \epsilon \m{E}^{\parallel\ast}_{i}\cdot\m{E}^\parallel_{j} - \Delta(\epsilon^{-1}) \m{D}^{\perp\ast}_{i}\cdot \m{D}^\perp_{j}\right)}{\sqrt{\int dA \rho |\m{Q}|^2 \int dA \epsilon |\m{E}_i|^2 \int dA \epsilon |\m{E}_j|^2}},\label{eqn:gij}
\eea
where $\m{E}$, $\m{D}$, $\m{Q}$ and $t$ are the optical mode electric field, optical mode displacement field, mechanical mode displacement field, and the thickness of the slab or thin-film.  For convenience, in what follows we denote the overlap integral in equation (\ref{eqn:gij}) as $\bra{\m E_i} \m Q \ket{\m E_j}$.  Since the optical supermodes are nearly degenerate, we replace $\omega_{i,j}$ with either $\omega_+$ or $\omega_-$ with little error, and following the notation used previously denote cross-modal coupling as $h = g_{+,-}= g_{-,+}$. 

For the optical and mechanical modes of a single L2 cavity (see Fig.~\ref{fig:cavities_and_waveguides}(a) and Fig.~\ref{fig:cavities_and_waveguides}(f)) the optomechanical self-coupling is numerically calculated using (\ref{eqn:gij}) to be $g=\bra{\m E} \m Q \ket{\m E}/2\pi = 489~\text{kHz}$.  Then using the symmetry selection rules in the overlap integrals, the only coupling terms involving the $\m Q_-$ mode are the cross-coupling terms $\bra{\m E_\pm} \m Q_- \ket{\m E_\mp}$.  These terms are calculated to be $\bra{\m E_+} \m Q_- \ket{\m E_-}=\bra{\m E_-} \m Q_- \ket{\m E_+} = \bra{\m E} \m Q \ket{\m E}/\sqrt{2}$ to good approximation. For the supermodes of our double L2 cavity system this yields a cross-coupling rate of $h/2\pi = 346~\text{kHz}$.

\subsection{Implementation of Waveguides}
A line-defect on an optomechanical crystal acts as a waveguide for light and sound~\cite{ref:Chutinan2000,ref:Johnson4,safavi-naeini10}. In principle then, the same waveguide may be used to shuttle both the photons and phonons around on an OMC microchip. However, due to the different properties of optical and acoustic excitations, in particular, their typically disparate quality factors (roughly $10^6$ for $200$~THz photons and $10^4$ for GHz phonons in silicon), the cavity loading requirements may be different for photons and phonons. As such, it is more convenient to implement two physically separate sets of waveguides, one for optics and the other for mechanics.  

Our chosen line defect for optical waveguiding, shown in Fig. \ref{fig:cavities_and_waveguides}(b), consists of a row of
removed holes, with the rows above and below shifted toward one another by $W$ such that the distance between the
centers of the snowflakes across the line defect is $\sqrt{3}a - 2W$.  Simulations of this line defect waveguide show
that there are no acoustic waveguide bands resonant with the localized mechanical L2 cavity modes of interest (see
Fig. \ref{fig:cavities_and_waveguides}(c)), and therefore that this waveguide will not load the mechanical part of the
L2 cavity system. Optically, this line defect has a single optical band crossing the frequencies of the localized
optical L2 cavity modes of interest (see Fig. \ref{fig:cavities_and_waveguides}(d)), providing the single-mode optical
waveguiding required for the PPT.

The line defect used for the acoustic waveguide consists of a row of holes which have been reduced in size, as shown in Fig. \ref{fig:cavities_and_waveguides}(g)). By shrinking the size of the central row of snowflake holes by $18\%$, a single-mode acoustic waveguide is formed which spectrally overlaps the localized mechanical resonances of the L2 cavity (see Fig. \ref{fig:cavities_and_waveguides}(h)), and allows for mechanical coupling to the PPT. This same line defect waveguide is reflective for optical photons in the band of localized optical resonances of the L2 cavity (see Fig. \ref{fig:cavities_and_waveguides}(i)), and thus is isolated from the optical part of the PPT.  Below we discuss how both the optical and mechanical waveguides may be used to load the PPT resonant cavity.  

\subsection{Cavity-Waveguide Coupling}

\begin{figure}[htbp]
\begin{center}
\includegraphics[width=15cm]{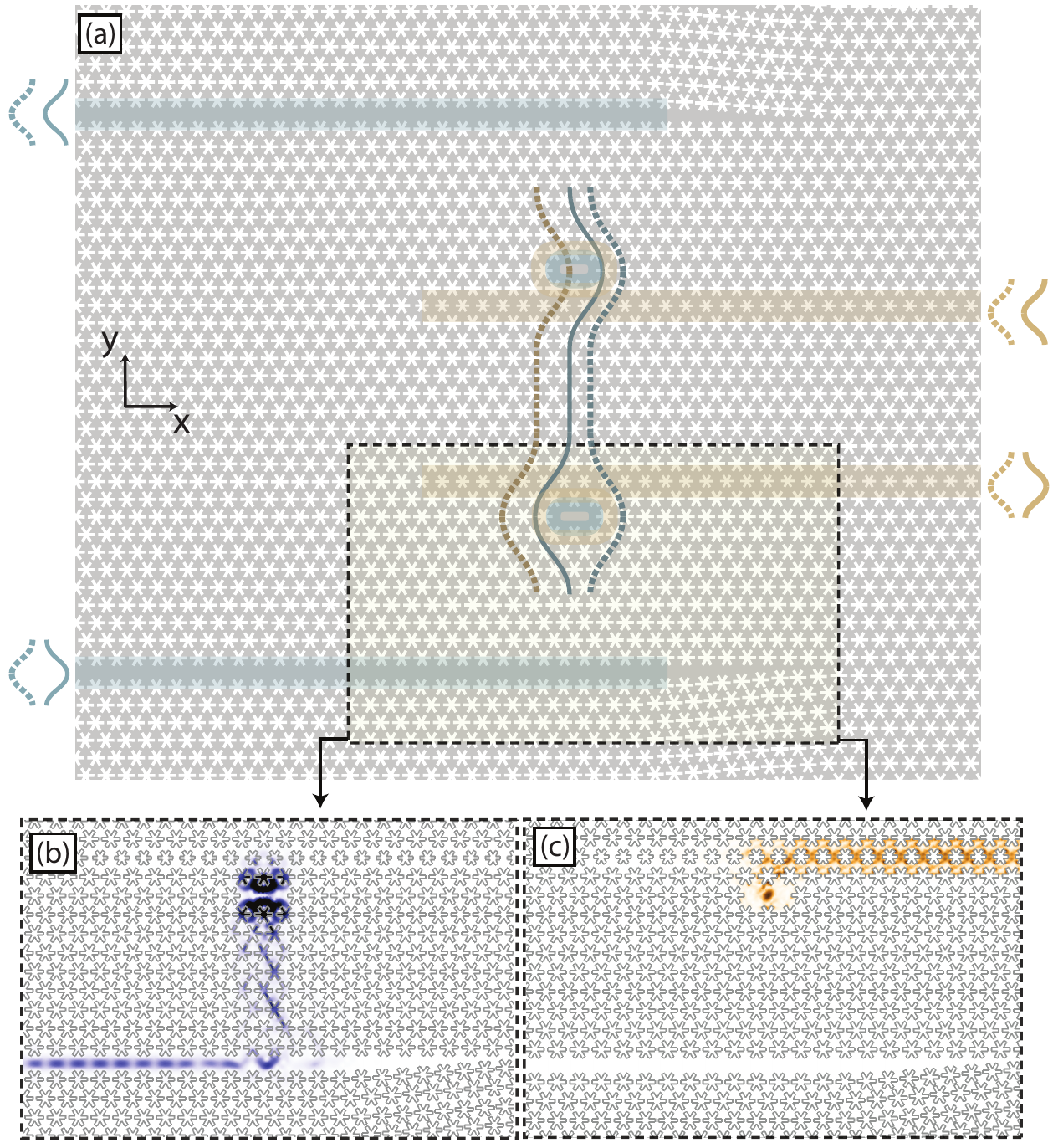}
\end{center}
\caption{(a) The full optomechanical crystal PPT system, consisting of a pair of coupled L2 defect cavities with acoustic and optical waveguide couplers.  The waveguide coupling to the cavities, for both optics and acoustics, consists of a pair of horizontal line defect waveguides, one to each of the L2 cavities.  The optical waveguides (highlighted in blue) are the two outer waveguides and the acoustic waveguides (highlighted in beige) are the two inner waveguides.  The relevant symmetric and anti-symmetric optical (blue) and acoustic (beige) supermodes of the cavity-waveguide system are shown as envelope functions with the appropriate symmetry (dashed for symmetric and solid for anti-symmetric).  The FEM-simulated (b) photonic ($|\avg{\m S_o}_t|^2$) and (c) phononic ($|\avg{\m S_m}_t|^2$) Poynting vectors of the lower cavity-waveguide structure, illustrating the selective optical loading of the lower waveguide and the selective mechanical loading of the upper waveguide on the cavity.\label{fig:full_system_only}}
\end{figure}

By bringing the optical waveguide near the L2 cavity, the optical cavity resonance is evanescently coupled to the guided modes of the line-defect, as shown in Fig.~\ref{fig:full_system_only}(b). Control over this coupling rate is achieved at a coarse level by changing the distance (number of unit cells) between the cavity and waveguide. For the structure considered here, a gap of 8 rows is sufficient to achieve a coupling rate $\kappa_e$ in the desired range. A fine tuning of the coupling rate is accomplished by adjusting the waveguide width parameter, with a value of $W = 0.135 a$ resulting in a loaded optical cavity $Q$-factor of $Q_\text{WG,o} \approx 3\times10^5$ (the corresponding external waveguide coupling rate is $\kappa_e/2\pi = 300~\text{MHz}$). Considering that intrinsic optical $Q$-factors as high as $3\times10^6$ have been achieved in microfabricated thin-film silicon photonic crystal cavities similar to the sort studied here~\cite{Takahashi2009}, the calculated optical waveguide loading should put such a cavity structure well into the over-coupled regime ($\kappa_e/\kappa_i \approx 10$).  A short section in which $W$ is tapered is used to close off the waveguide on one side.

The same design procedure for the acoustic waveguide results in an evanescently coupled waveguide at a distance of only one row from the L2 cavity. Since the acoustic line-defect waveguide does not support Bloch modes at the optical cavity frequency, no additional loss is calculated for the optical cavity resonance. In this geometry the mechanical cavity loading is calculated to be $Q_\text{WG,m} = 1.3\times10^3$, corresponding to an extrinsic coupling rate $\gamma_e/2\pi = 4.4~\text{MHz}$. Taking $Q_i \approx 10^4$ as an achievable intrinsic mechanical $Q$-factor (we need a citation to the Berkeley group here for Si GHz resonators), such a loading also puts the mechanical system in the over-coupling regime ($\gamma_e/\gamma_i \approx 10$).

Simulations of the above cavity-waveguide couplings are performed using FEM~\cite{COMSOL2009} with absorbing boundaries at the ends of the waveguide. The resulting time-average electromagnetic Poynting vector $|\avg{\m S_o}_t|^2 = |\m E \times \m H^\ast|/2$ of the optical field leaking from the L2 optical cavity resonance is plotted in Fig.~\ref{fig:full_system_only}(b), while the mechanical Poynting vector $|\avg{\m S_m}_t|^2= |- \m v \cdot \m T|^2$ ($\m v$ is the velocity field, and $\m T$ the stress tensor) of the acoustic waves radiating from the mechanical mode of the L2 cavity is shown in Fig.~\ref{fig:full_system_only}(c). It is readily apparent from these two plots that the coupling of the two different waveguides to the L2 cavity act as desired; the acoustic radiation is coupled only to the phononic waveguide, and the optical radiation is coupled only to the photonic waveguide.

In order to individually address and out-couple from the even and odd symmetry cavity resonances of double-L2-cavity structure used to form the PPT, a pair of waveguides is used for each of the optical and mechanical couplings.  As shown in the overall PPT design of Figure~~\ref{fig:full_system_only}(a), each of the L2 cavities has an acoustic and an optical waveguide coupled to them.  Excitation of a pair of waveguides either in or out of phase would thus allow for coupling to the symmetric or anti-symmetric supermodes, respectively, of the double-L2-cavity.  Similarly, spatial filtering (via an integrated directional coupler or waveguide filter for instance) of the output of a pair of waveguides would allow for the selective read-out of either the symmetric or anti-symmetric cavity modes.  One could in principle utilize spectral filtering to perform the selective mode coupling; however, with the narrowband nature of the optical and mechanical supermode splittings, spatially independent channels of excitation and read-out may be a preferred option.  

In summary, the OMC PPT as designed couples the symmetric and anti-symmetric optical modes of a double-L2-cavity system via a co-localized anti-symmetric mechanical resonance at frequency $\nu_m = 11.2~\text{GHz}$.  In the notation of Section~\ref{sec:analysis}, the lower frequency symmetric optical mode is the pump mode (cavity mode $a_{1}$), the anti-symmetric mode is the signal mode or cavity mode $a_{2}$, and the anti-symmetric mechanical resonance is phonon mode $b$.  Both optical modes are designed to have a resonant frequency in the near-IR around a frequency of $200$~THz, with a frequency splitting engineered to be equal to the mechanical frequency, $\omega_- - \omega_+ \approx\Omega=2\pi \times 11.2$~GHz.  The numerically calculated waveguide and optomechanical coupling rates for this system are $(\kappa_e, \gamma_e, h) = 2\pi\times( 300, 4.4, 0.35)~\text{MHz}$, with the required number of intracavity pump photons for optimum operation ($G\approx\sqrt{\kappa_{e}\gamma_{e}}$) of such a PPT estimated to be only $|\alpha_{1,0}|^2 = 1.1\times10^4$ (assuming minimal intrinsic losses and $\kappa_e\approx\kappa$, $\gamma_e\approx\gamma$). 

\section{Applications} \label{sec:applications}

At the simplest level, the extremely narrow optical response of the PPT, as shown in Figure~\ref{fig:scattering_matrix}, provides the opportunity for design and fabrication of filters with MHz-scale linewidths in the optical domain. By comparison, a purely passive optical design would require optical cavities with quality factors of $Q\approx10^8$.  More generally, such a scheme demonstrates a promising aspect of optomechanics in the realm of optical information processing. In this section three examples applications of the PPT are studied in detail. The first two, optical delay lines and wavelength converters are further examples of optical information processing that is of considerable interest in both classical and quantum information processing. The last application, using the PPT to provide optical ``flying qubit'' capability to superconducting microwave quantum systems, is an example of how optomechanics can have a fundamental role in hybrid quantum system engineering.

\subsection{Delay Lines}

\begin{figure}[htbp]
\begin{center}
\includegraphics{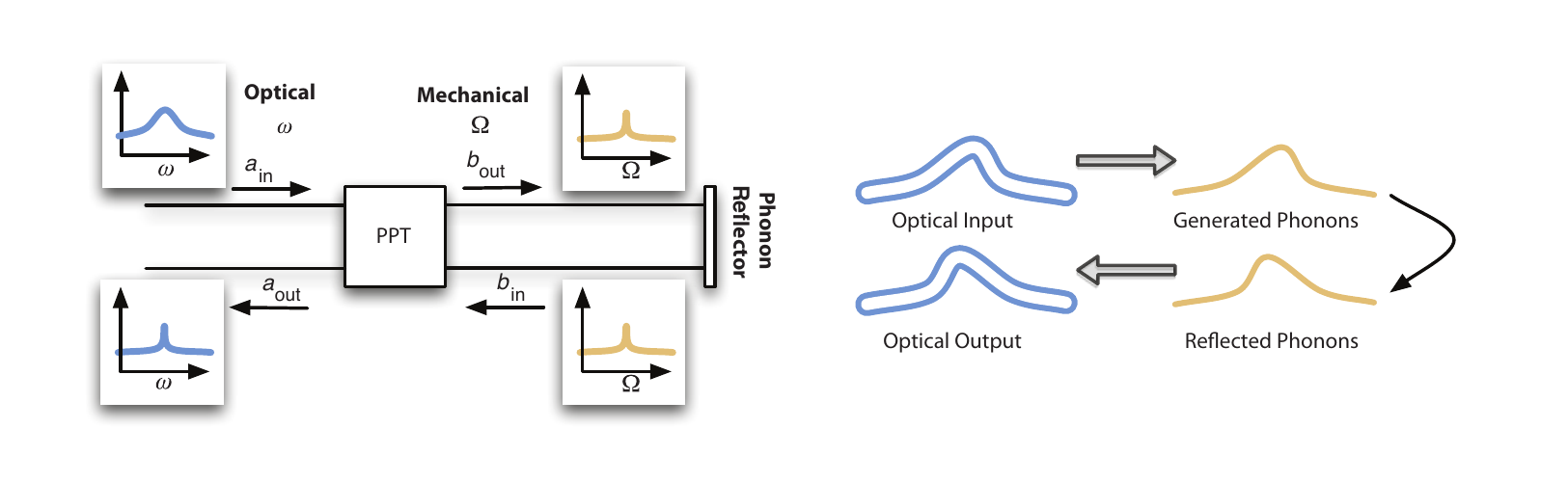}
\caption{Schematic layout (left) and signal path (right) of a PPT used in an optical delay line/filter configuration in which a phonon mirror at the end of the acoustic output reflects outgoing phonons back into the PPT.  The large optical delay is afforded by the acoustic path length of the signal in which acoustic waves propagate some $10^5$ times slower than photons.  Optical filtering is provided by the narrow resonance bandwidth of the mechanical component of the PPT.}
\label{fig:delay_line}
\end{center}
\end{figure}

Efficient reversible conversion between traveling photons and phonons can be used to realize an optical delay line, as shown schematically in Figure~\ref{fig:delay_line}.  In this geometry, the acoustic waveguide used to extract phonons from the PPT is terminated abruptly, forming an effective acoustic wave mirror which reflects phonons back toward the PPT after some propagation distance and delay.  Resonant photons sent into the optical port of the PPT, will then re-emerge, reflected and delayed by twice the length of the acoustic waveguide. The delay line functionality comes from the inherent slowness of acoustic waves in comparison to electromagnetic waves (roughly a factor of $10^5$ for waves in silicon). For a similar reason, electro-acoustic piezoelectric structures are used to create the chip-scale RF-microwave filters found in many compact wireless communication devices~\cite{Lakin1995}.  

The usefulness and bounds on the main characteristics of a PPT-based delay line, i.e., the total delay possible and the delay-bandwidth product, may be simply estimated without referring to a particular implementation of the system. The maximum possible delay is given by the lifetime of an excitation on the mechanical side of the system (cavity and waveguide), and is given by $1/\gamma_i$, which is limited by material properties of the mechanical system. The bandwidth of PPT conversion is given by eqn.~(\ref{eqn:gamma_transfer}), and is the total loss rate seen by the mechanical resonance, $2\gamma$. As such the delay-bandwidth product can at most be
\be
\Delta \omega \tau \sim \frac{2\gamma}{\gamma_i},
\ee
which is approximately twice the acoustic waveguide to mechanical resonance over-coupling ratio $\gamma_e/\gamma_i$ in the PPT. Manipulation of the acoustic waves within the delay waveguide itself, before conversion back to optics, may also be envisioned, enabling existing phononic information processing capabilities~\cite{Olsson2009} to be applied to optical signals.

\subsection{Wavelength Conversion}
\begin{figure}[htbp]
\begin{center}
\includegraphics{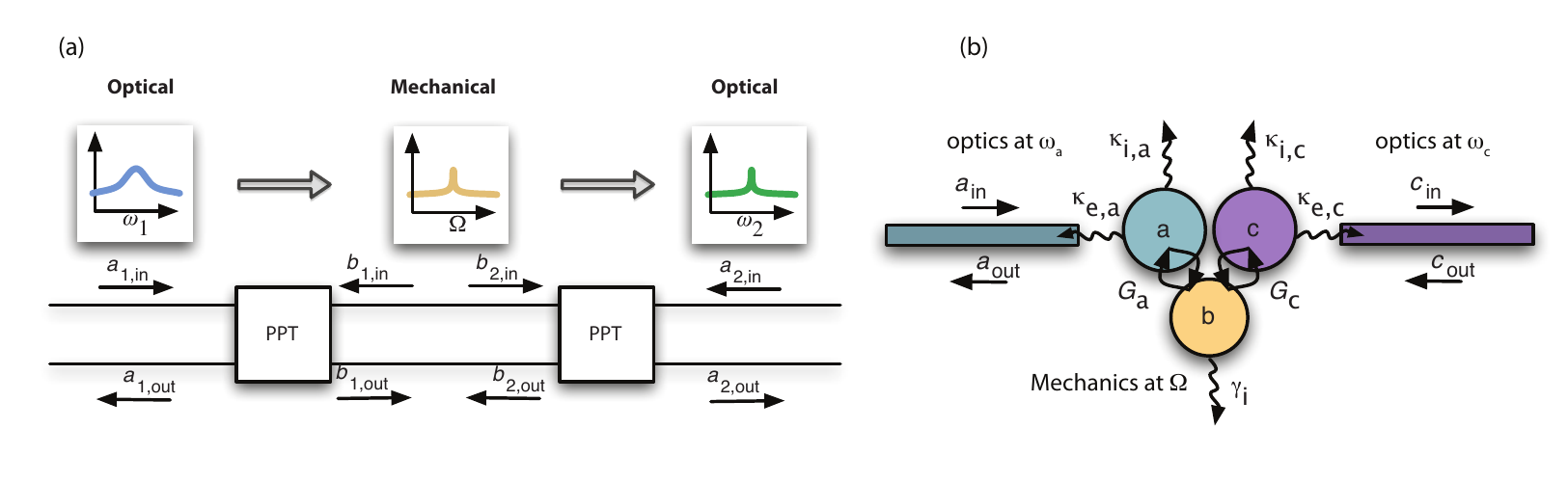}
\caption{(a) Schematic of a back-to-back PPT structure for optical wavelength convertsion/delay. For pure wavelength conversion, the phononic waveguide is vestigial, and can be removed by simply coupling the optical elements in both PPTs to the same mechanical resonant element. The simplified effective system diagram for such a device is shown in (b), where optical resonances $\hat{a}$ and $\hat{c}$ are coupled to the same mechanical resonance $\hat{b}$. The pump cavities for each system are omitted.}
\label{fig:ppt_wavelength_converter}
\end{center}
\end{figure}
Figure~\ref{fig:ppt_wavelength_converter}(a) shows the schematic for a PPT based device that can act simultaneously as a narrowband filter, a delay line and a wavelength converter.  It consists of connecting serially by a common acoustic waveguide, two PPT devices operating at different optical but matched mechanical frequencies.  Interestingly, if the only goal is to perform photon-to-photon wavelength conversion, one can omit the connecting acoustic waveguide. By placing the two PPTs ``on top of each other'', such that the optical cavities on both PPTs are coupled to the same mechanical resonance, as shown in Figure~\ref{fig:ppt_wavelength_converter}(b), optical wavelength conversion can be accomplished.  Such a PPT geometry could be realized by either having photonic cavities with multiple modes, or using two photonic cavities coupled to the same mechanical mode.  Such photon-to-photon conversion could even be taken to an extreme, allowing, for instance, optical-to-microwave wavelength conversion if one of the photonic cavities is a microwave cavity.

For the simplified wavelength conversion system of Fig.~\ref{fig:ppt_wavelength_converter}(b), the PPT matching condition of equation (\ref{eqn:matching_condition}) and noise analysis of Section~\ref{ss:noise} carry over with only minor adjustments. For the simplified wavelength conversion system the thermal noise is now split between the two optical channels ($\hat{a}$ and $\hat{c}$ of Fig.~\ref{fig:ppt_wavelength_converter}(b)), while the spontaneous emission noise in the system is approximately doubled (for similar optical cavities) due to the two uncorrelated spontaneous emission processes occurring from the optical pumping of each individual cavity.  In a single element PPT, the optimal $G$ matches the pure mechanical damping of the mechanical resonance ($\gamma$) to the induced optomechanical loading of the mechanical resonance ($G^2/\kappa$) by the optical cavity.  The matching condition for the simplified wavelength converter now must balance a mechanical resonance coupled on one side to an optical cavity with induced mechanical loading rate $G_a^2/\kappa_a$ and on the other side to a second optical cavity with induced loading rate $G_c^2/\kappa_c$.  As such, assuming that $\gamma_i \ll G_k^2/\kappa_k$ we arrive at the \emph{photon-photon converter} matching condition,
\be
\frac{G_a^2}{\kappa_a} = \frac{G_c^2}{\kappa_c}.
\ee
The optomechanical system as described would act as a quantum-limited optomechanical wavelength converter. Finally, we note that this particular implementation of the wavelength converter could also function in a wider array of optomechanical platforms since there is no longer a need for phononic waveguides.

\subsection{Quantum State Transfer and Networking between Circuit QED and Optics}

\begin{figure*}[htbp]
\begin{center}
\scalebox{0.70}{\includegraphics{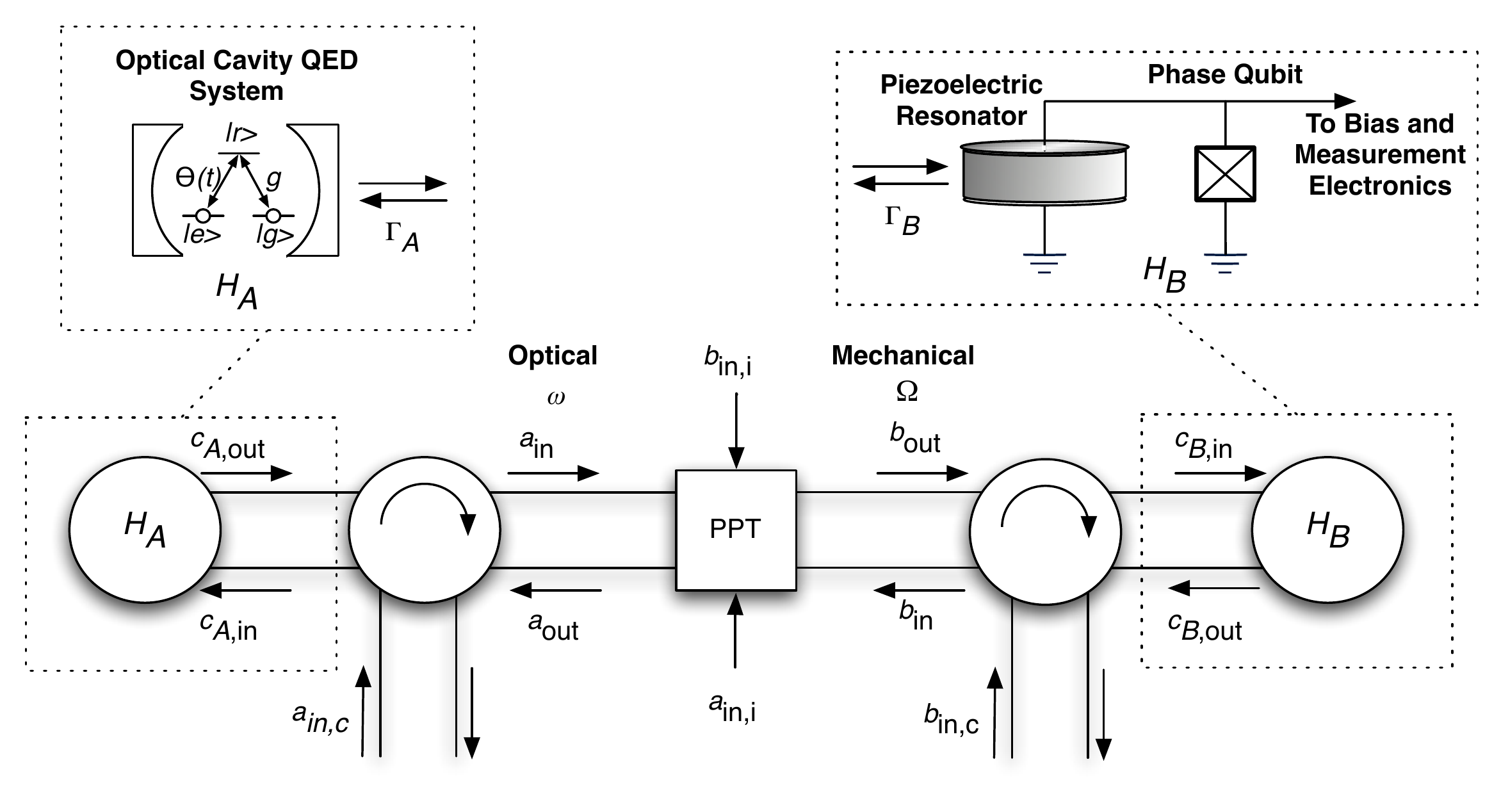}}
\caption{System diagram for quantum state transfer from optical to superconducting qubits. The two 3-port components are circulators with each input connected to the output directly clockwise from it. }
\label{fig:ciraczoller}  
\end{center}
\end{figure*}

Two of the key requirements for a viable platform for quantum computation are the ability to do storage and communication of quantum information. For the case of superconducting phase qubits, promising theoretical proposals to provide 
such functionality have involved electrical~\cite{Plastina2003} and mechanical resonators~\cite{Cleland2004,Geller2005}. Experimentally, an electromagnetic resonator quantum bus was demonstrated by Sillanp\"a\"a et al.~\cite{Sillanpaa2007}  in 2007, while more recently O'Connell et al.~\cite{OConnell2010} demonstrated the strong coupling of a mechanical resonator to a superconducting qubit. Circuit QED (cQED) to date remains limited by the lack of a true long range state transfer mechanism, one which is readily available for the case of quantum-optical qubits, in the form of optical fibers and free-space links. Using the PPT system, one potentially could implement a version of the quantum state transfer protocol of Cirac et al.~\cite{Cirac1997} allowing for the high fidelity transfer of states between optical and superconducting qubits. Such a system would satisfy one of the original goals of hybrid quantum system~\cite{Tian2004} by interfacing a quantum optical and solid-state qubit. 

By connecting the phononic waveguide of a PPT to a piezoelectric resonator strongly coupled to a superconducting qubit~\cite{Cleland2004,Geller2005,OConnell2010}, and connecting the PPT at its optical end to an optical cavity QED system, as shown in Figure~\ref{fig:ciraczoller}, the phonon-photon translator could be used as an intermediary in a state transfer protocol among two energy-disparate quantum systems. The optical system $(A)$ is composed of a Fabry-Perot cavity containing a three-level atom system in a $\Lambda$ configuration. As shown in~\cite{Cirac1997}, in the correct limit, this system may be modeled with effective Jaynes-Cummings (JC) Hamiltonian with a Rabi frequency $g_A(t)$ controlled by another beam. For the superconducting system $(B)$, a mechanical resonance is coupled to a phase qubit, with a bias current used to change the frequency of the resultant two-level system, which effectively changes the coupling rate between the qubit and mechanical resonator. This leads again to a system with an externally controllable Rabi frequency of $g_B(t)$. The Hamiltonian of each subsystem is then given by
\be 
\hat{H}_{j} =  \hbar g_j(t) e^{-i \phi_j(t)} \hat{\sigma}_j \hat{c}^\dagger_j + \hbar g_j(t) e^{+i \phi_j(t)} \hat{\sigma}^\dagger_j \hat{c}_j\qquad(j=A,B),
\ee
where $\hat{c}_j$ are the annihilation operators of the photonic or phononic resonances external to the PPT,  and $\hat{\sigma}_j$ are the level lowering operators for the respective qubits to which they are coupled.

Each cavity mode, with annihilation operators $\hat{c}_j$, is  coupled to its respective waveguide with a loss rate $\Gamma_j$. To characterize the PPT, the intrinsic losses in these cavities are ignored. Additionally, the phononic waveguide is assumed to be loss-less, though losses may be taken into account through the minor readjustment of the scattering parameters studied in Section~\ref{app:waveguide_loss}.  Using the input-output boundary conditions~\cite{Gardiner1993,GardinerZoller} the frequency domain expression for the noise input into systems $A$ and $B$ are found to be,
\begin{eqnarray}
\tilde{c}_{A,\text{in}}(\omega) &=&  \tilde{a}_\text{in,c}(\omega),\\
\tilde{c}_{B,\text{in}}(\omega) &=& s_{21}(\omega) \sqrt{\Gamma_A} \tilde{c}_A(\omega)  + s_{21}(\omega) \tilde{a}_\text{in,c}(\omega)
+ s_{22}(\omega) \tilde{b}_\text{in,c}(\omega)  \nonumber\\&&+ n_{12}(\omega) \tilde{a}_\text{in,i}(\omega)
+ n_{22}(\omega) \tilde{b}_\text{in,i}(\omega),
\end{eqnarray}
where $\hat{a}_{\text{in,c}}$ and $\hat{b}_{\text{in,c}}$ represent the noise being coupled into the system from the third input of each circulator in Figure~\ref{fig:ciraczoller}.  To convert this equation to the time-domain, the convolution between various operators and scattering matrix elements must be taken. If the photon pulse is of sufficiently large temporal width, i.e., with bandwidth less than the bandwidth of the PPT, the frequency dependence of each scattering matrix element can be removed, replacing it with its extremal value assuming that the system is operating at resonance $(\D = 0)$. This requires that the coupling rates $g_j(t)$ should change slowly relative to the response of the PPT. Under this condition, the input-output relations in the time-domain are then
\begin{eqnarray}
\hat{c}_{A,\text{in}}(t) &=& \hat{a}_\text{in,c}(t), \\
\hat{c}_{B,\text{in}}(t) &=& s_{21} \sqrt{\Gamma_A} \hat{c}_A(t)  + s_{21}\hat{a}_\text{in,c}(t) \nonumber\\&&+ s_{22} \hat{b}_\text{in,c}(t) + n_{12} \hat{a}_\text{in,i}(t)+ n_{22}\hat{b}_\text{in,i}(t).
\end{eqnarray}
In modeling the noise of the system, it is assumed that the optical noise inputs are in the vacuum state, and that the phononic noise is thermal with thermal phonon occupation numbers $\bar{n}$ and $\bar{n}^\prime$ for $\hat{b}_\text{in,c}(t)$  and $\hat{b}_\text{in,i}(t)$ , respectively, where the PPT phonon spontaneous emission noise is combined with the intrinsic thermal bath coupling of the mechanical mode in $\bar{n}^\prime$  as described in section~\ref{ss:spont_emission}.

Using standard operational methods of Quantum Stochastic Differential Equations~\cite{GardinerZoller,LesHouches1995}, the master equation describing the evolution of systems $A$ and $B$ is found to be, 
\begin{eqnarray}
\dot{\rho} &=& \frac{1}{i\hbar} [ H_A + H_B, \rho] + \frac{\Gamma_A}{2}\mathcal{L}_{I, A}\rho +\frac{\Gamma_B}{2}\mathcal{L}_{I, B}\rho \nonumber\\&&
+ \frac{\Gamma_B}{2}\left(|s_{22}|^2 \bar{n} + |n_{22}|^2 \bar{n}^\prime \right)\mathcal{L}_{T,B}\rho\nonumber\\&&
+\sqrt{\Gamma_A\Gamma_B}|s_{21}| \left([ c_B^\dagger,c_A \rho] + [\rho c_A^\dagger,c_B]\right)\label{eqn:AB_mastereqn}
\end{eqnarray}
with the Liouvillians $\mathcal{L}_{I, j}$ and $\mathcal{L}_{T, j}$ are given by,
\begin{eqnarray}
\mathcal{L}_{I, j}\rho &=&  2 \hat{c}_j \rho \hat{c}_j^\dagger -  \hat{c}_j^\dagger  \hat{c}_j \rho - \rho \hat{c}_j^\dagger  \hat{c}_j,\\
\mathcal{L}_{T, j}\rho &=&  \mathcal{L}_{I, A}\rho
+2 \hat{c}_j^\dagger \rho \hat{c}_j -  \hat{c}_j  \hat{c}_j^\dagger \rho - \rho \hat{c}_j  \hat{c}_j^\dagger.
\end{eqnarray}
\noindent The final term in the master equation~(\ref{eqn:AB_mastereqn}) is the cascading term \cite{Gardiner1993,LesHouches1995,Carmichael1993} which gives rise to the unidirectional coupling between the systems.

For the PPT, parameters typical to an OMC structure such as the one with scattering matrices plotted in Fig.~\ref{fig:scattering_matrix} are used, $(\gamma_e,\gamma_i,\kappa_e,\kappa_i,\Omega)=2\pi\times(10,1,2000,200,8000) ~\text{MHz}$. Equation~(\ref{eqn:Go_mod}) can be used to find the optimal matching optomechanical coupling rate, which for the assumed PPT parameters is $G^\text{o} = 155.4~\text{MHz}$.  At this optomechanical coupling rate, the resonant noise and scattering matrix parameters of the PPT are $(s_{21},s_{22},n_{21},n_{22}) = (0.917, 0.074, 0.290, 0.262)$, with a spontaneous emission noise equivalent occupation number of $n_\text{spon} = 0.200$.  Assuming that the PPT is cooled to the same cryogenic temperature of the superconducting qubit system ($T<100$~mK), the thermal bath component of the effective thermal occupancy of the PPT mechanical resonance can be neglected, and $\bar{n}^\prime \approx 0.251$.

The exact functional form of the $g_j(t)$ are found through numerical optimization.  This was done by taking the pulse-shape to be a step smoothed by a sinusoidal function with a rise (fall) time of $t_{r}$ ($t_{f}$). Optimization on the state transfer fidelity for an ideal PPT ($s_{21} = 1$) and for circulators running in the direction shown in Fig.~\ref{fig:ciraczoller} (qubit transfer from optical to superconducting system) leads to rise and fall times of $t_A = 23.5~\mu\text{s}$ and $t_B = 16~\mu\text{s}$, which are within the modeled PPT's $(2\gamma)/2\pi=22~\text{MHz}$ bandwidth.

\begin{figure}[htb]
\begin{center}
\scalebox{0.8}{\includegraphics{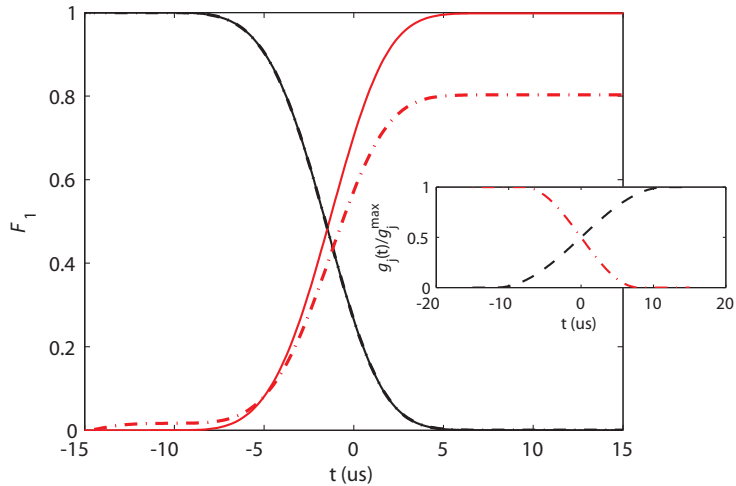}}
\caption{The main plot shows the time variation of the fidelity $F_1$ found by numerically solving the quantum master equation for a version of the quantum state transfer protocol introduced by Cirac, et al.~\cite{Cirac1997} for the state transfer from the optical to superconducting qubit.  PPT parameters of $(\gamma_e,\gamma_i,\kappa_e,\kappa_i,\Omega)=2\pi\times(10,1,2000,200,8000) ~\text{MHz}$ are used here to find the the optical ($F^{B,A}_1$; \textcolor{black}{$-\cdot$}) and superconducting ($F^{A,B}_1$; \textcolor{red}{$-\cdot$}) state-transfer fidelity curves for an initial state of $|1\rangle_{A}|0\rangle_{B}$. We see that a maximum fidelity $F^{A,B}_1$ of approximately 0.8 for the final state transfer is possible. The continuous lines (\textcolor{black}{$-$}) and (\textcolor{red}{$-$}) were found by quantum simulation of an ideal PPT, i.e. $\gamma_i=\kappa_i = 0$. In all cases analyzed the external system parameters were $(\Gamma_A,\Gamma_B,g^{\text{max}}_A,g^{\text{max}}_B) = 2\pi\times(50,5.0,5.0,1.0) ~\text{MHz}$.  The inset plot shows the shape of the control pulse used for the optical (\textcolor{black}{$--$}) and superconducting (\textcolor{red}{$-\cdot$}) qubits.}
\label{fig:fidelity_plot}
\end{center}
\end{figure}
 
Putting this all together, in Fig.~\ref{fig:fidelity_plot} we plot estimates of the fidelity of the quantum state transfer between system $A$ and $B$ via the connecting PPT.  The definition of fidelity used to calculate the state transfer efficiency is $F^{A,B}_j=\text{Tr}_A(|j_B\rangle\langle j_B|\rho)$, where $j=0,1,+$ represent respectively the ground $|0\rangle$, excited $|1\rangle$, and $|+\rangle =2^{-1/2}(|0\rangle+|1\rangle)$ states of the atomic and superconducting two-level systems. Under these conditions, and considering an ideal PPT ($s_{21} = 1$), states are transfered with fidelities $F^{A,B}_1=0.9983$, $F^{A,B}_+=0.9995$ and $F^{A,B}_0 = 1.00$. Taking into account the actual scattering and noise matrix values given above for the PPT, and accounting for the spontaneous emission noise of the PPT we find that the fidelities are reduced to $F^{A,B}_1=0.803$, $F^{A,B}_+=0.936$ and $F^{A,B}_0 = 0.983$. The inverse system, with circulators turning the opposite direction to transfer qubits from the superconducting to optical system was also studied, for which the same input pulses only time reversed, and yield fidelities $F^{B,A}_1 = 0.772$, $F^{B,A}_+ = 0.904$ and $F^{B,A}_0 = 0.983$.

\section{Summary}

We have introduced the concept and design of a traveling phonon-photon translator. We have shown that with a realistic set of parameters and the use of existing silicon optomechanical crystal technology, efficient and reversible conversion between phonons and photons should be possible. By characterizing the noise processes experienced by such a device, both classically and quantum mechanically, we have shown the utility of traveling phonon-photon translation to important problems in both classical optical communication and quantum information processing.

\section{Acknowledgments}

The authors wish to thank Darrick Chang, Thiago Alegre, Matt Eichenfield, Klemens Hammerer, and Peter Zoller for useful discussions.  This work was supported by the DARPA/MTO ORCHID program through a grant from AFOSR, and the NSF through EMT grant no. 0622246 and CIAN grant no. EEC-0812072.  ASN acknowledges support through NSERC of Canada.

\section*{References}


\end{document}